\documentclass[%
twocolumn,
amstext,
amsfonts,
amsmath,
amssymb,
superscriptaddress,
floatfix,
showpacs,
nobibnotes,
nolongbibliography,
nofootinbib,
aps,
prd
]
{revtex4-2}

\usepackage[utf8]{inputenc}
\usepackage{xargs}
\usepackage[german, english]{babel}
\usepackage{graphicx}
\usepackage{color}
\usepackage{cancel}
\usepackage{lineno}
\usepackage{xspace}
\usepackage{dcolumn}
\usepackage{ulem}
\usepackage{url}
\usepackage{bbm}
\usepackage{bm}
\usepackage{bbold}
\usepackage{float}
\usepackage{subfigure}
\usepackage{rotating}
\usepackage{siunitx}
\usepackage{mathrsfs}
\usepackage{slashed}
\usepackage{tensor}
\usepackage{nicefrac}
\usepackage{svn-multi}
\usepackage{enumitem}
\usepackage{wrapfig}
\usepackage{xcolor}

\usepackage{hyperref}
\hypersetup{
colorlinks=true,
citecolor=blue,
citebordercolor=red,
linktoc=all,
linkcolor=blue,
urlcolor=blue
}

\usepackage{natbib,times}
\newcommand{\onefig}{0.49\textwidth}
\newcommand{\twofigs}{0.49\textwidth}

\newcommand{\E}{\mathrm{e}}

\newcommand{\Tr}{\mathrm{Tr}}
\newcommand{\tr}{\mathrm{tr}}

\newcommand{\vev}[1]{\ensuremath{\left\langle #1 \right\rangle}}

\def\Fig#1{Fig.~\ref{#1}}

\renewcommand*{\eqref}[1]{Eq.~(\ref{#1})}

\def\roughly#1{\mathrel{\raise.3ex\hbox{$#1$\kern-.75em%
\lower1ex\hbox{$\sim$}}}}

\usepackage{titlesec}
\titleformat*{\paragraph}{\normalfont}{}{}{}

\graphicspath{
{./}
{./figures/}
}

\newcommand{\JLU}{%
	Institut f\"{u}r Theoretische Physik, %
	Justus-Liebig-Universit\"{a}t Gie\ss{}en, %
	35392 Gie\ss{}en, %
	Germany%
}

\newcommand{\HFHF}{%
	Helmholtz Forschungsakademie Hessen f\"{u}r FAIR (HFHF), %
	GSI Helmholtzzentrum f\"{u}r Schwerionenforschung, %
	Campus Gie\ss{}en, %
	35392 Gie\ss{}en, %
	Germany%
}

\begin{document}

\title{Regulator scheme dependence of the chiral phase
  transition at high densities}

\author{Konstantin Otto}
\email[E-Mail:]{konstantin.otto@physik.uni-giessen.de}
\affiliation{\JLU\\[0.5ex]} 
\affiliation{\HFHF}
\author{Christopher Busch}
\email[E-Mail:]{christopher.busch@physik.uni-giessen.de}
\affiliation{\JLU\\[0.5ex]}
\affiliation{\HFHF}
\author{Bernd-Jochen Schaefer}
\email[E-Mail:]{bernd-jochen.schaefer@theo.physik.uni-giessen.de}
\affiliation{\JLU\\[0.5ex]} 
\affiliation{\HFHF}

\pacs{
12.38.Aw, 
11.30.Rd, 
11.10.Wx, 
05.10.Cc 
}

\begin{abstract}
  A common feature of recent functional renormalization group
  investigations of effective low-energy QCD is the appearance of a
  back-bending behavior of the chiral phase transition line at low
  temperatures together with a negative entropy density in the
  symmetric regime. The regulator scheme dependence of this phenomenon
  and the necessary modifications at finite densities are analyzed
  within a two-flavor quark-meson model. The flows at finite densities
  for three different regulators of three- or four-dimensional momenta
  are confronted to each other. It is found that the back-bending
  behavior and the negative entropy density can be traced back to the
  explicit momentum dependence of the regulator shape function. While
  it persists for the often-used three-dimensional flat regulator, it
  vanishes for Callan-Symanzik type regulators. This
    points to truncation artifacts in the lowest order of the
    derivative expansion.  A careful theoretical as well as numerical
  exploration is given.
\end{abstract}

\maketitle

\section{Introduction}
\label{sec:introduction}

Until now, the QCD phase structure at high densities and low
temperatures is mostly an unknown territory. Several distinct phases
are expected to exist in this regime making its phase structure
extremely rich \cite{Bzdak:2020bjaa}. Examples of cold and dense
strongly-interacting matter cover exotic phases of QCD such as
crystalline color-superconductor, 2SC and color flavor locked (CFL)
phases and even possibly spatially inhomogeneous phases, for reviews
see e.g.~\cite{Alford:2007xm, Buballa:2014tba}.  Most of these phases
are difficult if possible at all to achieve in laboratory experiments.
Available experimental data in this area of the phase diagram is
still limited and has rather poor statistics, cf.~the recent beam
energy scans (BES) at RHIC in Brookhaven \cite{Tlusty:2018rif}.
Several upcoming experimental facilities such as the CBM
\cite{Klochkov:2021eyo} or NICA \cite{Butenko:2021wfr} experiments
were designed to fill the gap and to explore this region with higher
statistics in the near future.

From a theoretical point of view, state-of-the-art lattice simulations
at finite chemical potentials are hampered by a sign problem
\cite{deForcrand:2010ys} such that alternative approaches are
necessary to investigate the intermediate-density region of the phase
diagram.  Most widely utilized tools in this context are effective
models that by sharing some important symmetries with QCD are expected
to reflect some of its characteristic properties. A notable example is
the quark-meson model, also known as a linear sigma model
combined with quarks, wherein the effective low-energy couplings arise
from the integration of gluonic degrees of freedom \cite{Jungnickel1995,Schaefer:2006sr, Fu:2021oaw}. Despite the crude
simplification compared to full QCD these models do incorporate
important phenomena such as the spontaneous chiral symmetry breaking
and its restoration at finite temperature and/or density. However,
most often such effective model approaches are typically studied in
mean-field approximations where important non-perturbative quantum and
thermal fluctuations are neglected, though they are of utmost
relevance in particular in the vicinity of any phase transitions. The
situation can be much improved by combining these models with
functional methods
like the functional renormalization group (FRG), Dyson-Schwinger
equations (DSEs) and $n$-PI approaches, thus making contact with the
underlying full QCD.  These methods are an essential and very powerful
framework in the study of such non-perturbative issues.  Recent
elaborate studies in the context of full QCD with the FRG
\cite{Fu:2019hdw} and the DSE \cite{Isserstedt:2019pgx,
  Fischer:2018sdj} or even combinations of both \cite{Gao:2020qsj,
  Gao:2020fbl} suggest the existence of a critical endpoint at
intermediate chemical potentials and temperatures, implying a chiral
phase transition from a crossover to a first-order transition at
increasing density. However, in the high-density region the physics
becomes more involved which makes the needed and necessary truncations
in the functional approaches rapidly inapplicable. Due to the lack of
feasible first-principle computations in this region one is mainly led
to the reliance on some simplified truncations within the functional
framework combined with effective theories so far \cite{Braun:2022olp,
  Braun:2021uua}.

The spontaneous breaking of chiral symmetry and its restoration at
finite density has been studied extensively with the FRG by several
groups in the past, for a recent review see \cite{Fu:2022gou}.  First
applications also to neutron star physics have been made only recently
\cite{Drews:2014spa, Otto:2019zjy, Otto:2020hoz, Leonhardt:2019fua}. A
still open question, however, is posed by the back-bending behavior of
the chiral transition line at smaller temperatures
\cite{Schaefer:2004en} and the simultaneous occurrence of a negative
entropy densities beyond the chiral transition as firstly discussed in
\cite{Tripolt2017}.  Therein, it was speculated that besides a
truncation or scheme dependent artifact, this phenomenon could also be
related to an incorrect assumed vacuum state of the FRG method, caused
by, e.g., the formation of diquark condensates or the existence of
inhomogeneous phases. The back-bending of the chiral transition line
has also been found in other models \cite{Tripolt:2021jtp,
  Braun:2017srn} as well as with different numerical solution methods
such as the discontinuous Galerkin method \cite{Grossi:2021ksl,
  Grossi:2019urj} or the global pseudospectral Chebyshev expansion
method \cite{Chen:2021iuo}.  All these findings exclude numerical
solution artefacts impressively. Hence, the back-bending behavior and
the appearance of the negative entropy density seems to be reasoned in
the structure of the corresponding flow equations.  In this work, we
will attempt to solve this riddle by investigating the regulator
scheme dependence of the high-density chiral phase transition.

The paper is organized as follows: After a brief recapitulation of the
employed functional renormalization group method in Sec. \ref{sec:FRG}
the inherent regulator scheme dependence of the flow equations for
four-dimensional quantum field theories in local potential
approximation are addressed in Sec. \ref{sec:regulator-functions}. The
necessary modifications of the regulator at finite densities are
elaborated in the following and the relation to the Silver Blaze
property is summarized in a general framework.  In the next
Sec.~\ref{sec:qm_model} a quark-meson model truncation for two flavors
is introduced and the flow equations for three different regulator
choices as well as a parameter fixing procedure are outlined.  In
Sec.~\ref{sec:numerical_results} the regulator scheme dependence of
the phase structure in local potential approximation is analyzed. The
back-bending of the chiral phase transition and its relation to the
choice of the regulator is elucidated. We end with a detailed analysis
of the decoupling of the fermions from the flow and conclude in
Sec.~\ref{sec:summary}.  A discussion of regulator optimization and
further details such as the employed numerical methods and a
discussion on the pole proximity of the vacuum flows are collected in
four appendices \ref{app:optimization}-\ref{app:numimpl}.

\section{Functional Renormalization Group}
\label{sec:FRG}

In order to make this work self-contained, we briefly recapitulate
here important ingredients of the FRG which are needed for the present
analysis. For QCD-related recent reviews see
  Refs.~\cite{Pawlowski:2005xe, Gies2015, Schaefer:2006sr,
    Braun:2011pp, Fu:2022gou}, for a recent global review on physics applications of the FRG
including QCD see~\cite{Dupuis:2020fhh}.
As already mentioned the FRG is a suitable non-perturbative approach
towards solving continuum quantum field theories.  A modern
realization of Wilson's RG idea in terms of a functional differential
equation for the 1PI effective average action $\Gamma_k$
\cite{Ellwanger:1993mw,Morris:1993qb} is known as
the Wetterich equation \cite{Wetterich:1992yh}
\begin{equation}
\label{eq:Wetterich_eq}
\partial_t \Gamma_k [\phi]= \frac{1}{2} \Tr \left[ \partial_t R_k
  \left(\Gamma_k^{(2)} [\phi]+R_k\right)^{-1} \right] \ . 
\end{equation}
Here, $\Gamma_k^{(2)} [\phi]$ denotes the second functional derivative
with respect to the given field $\phi$ and $t=\ln (k/\Lambda)$ is the
logarithmic RG scale relative to an initial UV momentum cutoff scale
$\Lambda$.  The trace runs over the momenta and all inner spaces such
as flavor, spin or Dirac space.

\eqref{eq:Wetterich_eq} is a functional partial differential equation
for $\Gamma_k$ with a one-loop structure and interpolates between the
microscopic bare action $S$ in the UV and the full macroscopic quantum
effective action $\Gamma = \Gamma_{k=0}$ in the infrared (IR).  An
important ingredient of \eqref{eq:Wetterich_eq} is the momentum
regulator $R_k$ in the inverse propagator.  For real bosonic fields
this quantity is introduced in the action
$\sim \int_p \phi(-p) R_k(p^2) \phi (p)$ and refers to the regulator
scheme of the flow equation.\footnote{In the following we employ the
  short-hand momentum integration notation
  $\int_{p} \equiv \displaystyle \int \frac{\mathrm{d}^d p}{(2\pi)^d}$
  wherein the dimension $d$ is fixed by the one of index $p$.} It
introduces the scale parameter $k$ which describes the RG
coarse-graining.  The regulator has to fulfill some essential
properties which are crucial for this work which is why its
detailed discussion is postponed to the next section.  Basically, it
acts as an additional mass term to the low-momentum modes while the
insertion term in the momentum loop, the scale derivative of the
regulator $\partial_t R_k$, regularizes the ultraviolet modes, thus making
the flow IR and UV finite. This satisfies the RG notion of
successively integrating out quantum fluctuations in a shell around
the momentum $p \sim k$.  While the Wetterich equation is an exact
functional equation, in practice any attempt at its solution relies on
a truncation of the underlying functional $\Gamma_k$. One possible
truncation scheme is an expansion of $\Gamma_k$ in powers of
derivatives in four-dimensional configuration space, which e.g. for a
scalar theory with a real field $\phi(x)$ reads
\begin{equation}
  \label{eq:1}
\Gamma_k[\phi] = \int d^4 x \left[ V_k(\phi) + \frac{1}{2} Z_k(\phi)
  \left(\partial_\mu \phi \right)^2 +
  \mathcal{O}\left(\partial^4\right) \right] \ . 
\end{equation}
The leading order derivative expansion with a vanishing anomalous
dimension $Z_k \equiv 1$ provides the ansatz for the local potential
approximation (LPA). Then, for a constant vacuum expectation value
(VEV) $\vev{\phi} := \phi_0$, the two-point function always exhibits
the canonical momentum form $\Gamma_k^{(2)} = p^2 + m_k^2$ with the
effective momentum-independent curvature mass
$m_k^2 = \mathrm{d}^2 V_k(\phi)/\mathrm{d}\phi^2 |_{\phi=\phi_0}$.  In
general, the scale dependent effective potential $V_k$ assembles all
momentum independent field fluctuations to infinite order and thereby
dynamically modifies $m_k^2$.

\section{Regulator Schemes}
\label{sec:regulator-functions}

Any truncation of the effective average action necessarily leads to a
corresponding error. The possible types and sizes of such errors are
influenced among others by the choice of a suitable regulator function
so that this is an important ingredient in the FRG framework
\cite{DePolsi:2022wyb, Canet:2002gs}.

We focus our investigation on four-dimensional, local relativistic
quantum field theories and start with bosonic fields without chemical
potential. In momentum space, the regulator $R_k (p^2)$ has
squared-mass dimension and depends on a single momentum argument.
Any suitable regulator $R_k(p^2)$ can in principle be chosen at will
but should obey the following three restrictions:
\begin{align}
\begin{split}
  \label{eq:17}
  1. \quad &  \lim\limits_{k^2/p^2 \rightarrow 0} R_k(p^2) = 0\\
  2. \quad &  \lim\limits_{p^2/k^2 \rightarrow 0} R_k(p^2) > 0 \\
  3. \quad & \lim\limits_{k \rightarrow \infty} R_k(p^2)
  \rightarrow \infty \ .
\end{split}
\end{align}

The first requirement reflects the RG property that high momentum
modes are fully integrated out in the infrared. The vanishing of the
regulator ensures the crossing of the coarse-grained $\Gamma_k$ to the
full quantum 1PI effective action $\Gamma$. As a side remark,
$R_k(p^2)$ should vanish sufficiently fast,
$\lim\limits_{p^2/k^2 \rightarrow \infty} (p^2)^{(d-1)/2} R_k(p^2) =
0$, to obtain finite loop integrals in $d$
  dimensions.

The second requirement can be seen as an IR regularization such that
the effective propagator in LPA at vanishing field
$\Delta_k (p^2)= 1/(p^2+R_k (p^2))$ remains finite for $p^2 \to 0$, therefore
avoiding infrared divergences in the presence of massless modes.

The last requirement in \eqref{eq:17} just fixes the classical (bare)
action $S$ in the UV, e.g., for an effective theory
  with a finite UV cutoff $\Lambda$.

For convenience, a regulator function $R_k (p^2)$ that does not depend
on additional parameters can be rewritten in terms of a dimensionless
shape function $r(y)$
\begin{equation}
\label{eq:15}
R_k (p^2) = p^2 r(y) \qquad \text{with}\quad y=p^2/k^2
\end{equation}
by means of the (massless) dimensionless inverse propagator
\begin{equation}
\label{eq:inverse_propagator}
P^2(y) \equiv \frac{1}{k^2}  \Delta^{-1}_k(p^2) = y [1 + r(y)] \ .
\end{equation}

To minimize truncation errors, general optimization criteria for
regulator functions have been developed and are briefly reviewed in
App.~\ref{app:optimization}. They are typically designed to minimize
the regulator dependency of physical observables in the infrared and
explicit optimizations have so far only been conducted for flows at
vanishing chemical potential.

Beyond these criteria further requirements can be necessary.  Examples
are the preservation of the Silver Blaze property at finite chemical
potential (see next Sec. \ref{sec:silver_blaze}) or the Slavnov-Taylor
identities in gauge field theories, see e.g. \cite{Dupuis:2020fhh}.

Furthermore, we will argue in this work that there are additional
large regulator dependent truncation artifacts at small temperatures
and nonzero chemical potentials which lead to a back-bending
phenomenon of the chiral transition line and the corresponding
occurrence of negative entropy densities.

\subsection{Regulators and the Silver Blaze constraint}
\label{sec:silver_blaze}

At nonvanishing densities, we extend the definition of the regulator
to include a possible dependence on the associated chemical
potentials. This is necessary in general to ensure certain physical
properties of the theory.  An example is QCD with one chemical
potential as an external parameter.  For vanishing temperature this
yields a certain characteristic of the $n$-point functions
$\Gamma^{(n)}$ often dubbed as the \textsl{Silver Blaze} property in
the literature \cite{Cohen:2003kd}.  It states for a fixed vacuum
state and $\mu$ smaller than some critical chemical potential $\mu_c$
that the free energy of, e.g., a fermionic system
  does not exhibit a $\mu$-dependence at zero temperature. The
  critical chemical potential is determined by the pole mass of the
  lightest particle $\mu_c = m_{\mathrm{pole}}$ carrying a finite
  charge associated with the corresponding chemical potential.  This
  transfers to the correlation functions such that the
  $\mu$-dependence of $\Gamma^{(n)}$ is solely given by replacing the
  zero components of the four-momenta in the vacuum correlation
  functions with $\mu$-shifted zero components.  This becomes trivial
  for, e.g., a free Dirac fermion with mass $m$: the inverse
propagator with a chemical potential $\mu$ can be rewritten as
\begin{equation}
  \Gamma^{(2)}(p_1,p_2;\mu) = \frac{\delta(p_1-p_2)}{\mathrm{i}
    \slashed{p}_1 - \mu \gamma_0 + m} =
  \Gamma^{(2)}(\tilde{p}_1,\tilde{p}_2;0) 
\end{equation}
where $\tilde{p}_i := (p_i^0 + \mathrm{i} \mu, \bm{p}_i)$ denote the
shifted momenta.  For higher $n$-point functions and $\mu < \mu_c$
Silver Blaze generalizes to
\begin{equation}
\label{eq:silver_blaze}
\Gamma^{(n)}(p_1, \dots, p_n; \mu) = \Gamma^{(n)}(\tilde{p}_1, \dots,
\tilde{p}_n; 0)  
\end{equation}
with $\tilde{p}_i = (p_i^0 + \mathrm{i} c_i \mu,\bm{p}_i)$. The value
of $c_i$ determines how the corresponding field couples to the
chemical potential.  For example, augmenting the free Dirac theory
with bosons via, e.g., a Yukawa interaction, one has $c_i=1$ for
fermionic momenta and $c_i=0$ for bosonic ones.  See
\cite{Marko:2014hea, Braun:2017srn} for more details
and a proof of the Silver Blaze property in the functional 2PI
framework.

The Silver Blaze constraint is necessary for a consistent
thermodynamic treatment in particular close to a phase transition at
low temperatures. In the context of the FRG the preservation of
\eqref{eq:silver_blaze} in the infrared can be ensured by extending it
to all scales $k$
\begin{equation}
\label{eq:silver_blaze_k}
\Gamma^{(n)}_k(p_1, \dots, p_n; \mu) = \Gamma_k^{(n)}(\tilde{p}_1,
\dots, \tilde{p}_n; 0)
\end{equation}
where the threshold
  $\mu_{c,k} = m_{\mathrm{pole},k}$ is now scale-dependent since it is
  determined by the running pole mass \cite{Khan:2015puu, Fu:2016tey}.

This translates to a similar property of the fermionic regulator such
that it becomes $\mu$-dependent:
\begin{equation}
\label{eq:regulator_condition}
R_k^F(p;\mu) = R_k^F(\tilde{p};0)\ .
\end{equation}
As a consequence, in any loop diagram the frequency component of the
loop momentum can be shifted by $-\mathrm{i}\mu$ and the contour in
the complex plane can be closed. If no poles exist inside the closed
contour and all external momenta of the loop diagrams are also shifted
the vacuum result will be recovered.

For example, for a free Dirac field the zero temperature flow equation
for the effective fermionic potential $U_k^F$ (i.e., the negative
pressure) at finite $\mu$ is given by the loop integral
\begin{align}
\begin{split}
\label{eq:fermi_flow_sb}
\partial_t U_k^F &= - \tr \int\limits_{-\infty}^{\infty}\!\!
\frac{dp_0}{2\pi} \int\!\! \frac{d^3p}{(2\pi)^3} \frac{\partial_t
  R_k^F(p;\mu)}{\mathrm{i} \slashed{\tilde{p}}+m +
  R_k^F(p;\mu)}\\
&= - \tr\!\!\!\! \int\limits_{-\infty +\mathrm{i}\mu}^{\infty +\mathrm{i}\mu}\!\!\!\!
\frac{dp_0}{2\pi} \int\!\! \frac{d^3p}{(2\pi)^3} \frac{\partial_t
  R_k^F({p};0)}{\mathrm{i} \slashed{{p}}+m + R_k^F({p};0)}
\end{split}
\end{align}
which demonstrates the complex momentum shift compared to the original
vacuum flow.

While \eqref{eq:regulator_condition} poses a necessary condition for
the fulfillment of the Silver Blaze property, it is not sufficient.
To ensure a completely $\mu$-independent flow for $\mu < \mu_c$ the
running threshold $\mu_{c,k}$ must always be larger $\mu_c$ such that
the pole mass is approached from above in the infrared. This is
actually a challenging restriction for the regulator since additional
poles in the complex plane might be generated by the regulator
\cite{Floerchinger:2011sc}. An example is given by the exponential
regulator $r^{\text{exp}}(y) = 1/(\exp(y)-1)$ which generates
infinitely many complex propagator poles in the complex frequency
plane.

Explicit computations with regulators fulfilling all those conditions
typically constitute an arduous analytical and numerical task; see
e.g. \cite{Pawlowski:2015mia, Helmboldt:2014iya} for applications. A
simple way to circumvent these problems is to use dimensionally
reduced (i.e., purely spatial) regulators. In four
dimensions, a popular choice is a three-dimensional cutoff function
regularizing only the spatial modes:
\begin{equation}
\label{eq:3d_regulator}
R_k^\mathrm{3d}(\bm{p}^2) = \bm{p}^2 \, r(x) \ , \quad
R_k^{F,\mathrm{3d}}(\bm{p}) = \mathrm{i}\slashed{\bm{p}} \, r^F(x) 
\end{equation}
with $x:=\bm{p}^2/k^2$. Any $\mu$-dependence vanishes due to the
absence of the frequency argument.  Theories in the presence of such
regulators always retain the Silver Blaze property because
\eqref{eq:regulator_condition} is trivially fulfilled. For such a
regulator, the flow equation for the potential
(\ref{eq:fermi_flow_sb}) becomes
\begin{align}
  \partial_t U_k^F =- 4 \int_{\bm{p}} \frac{\bm{p}^2 \, ( 1+r^F(x) ) \, \partial_t
                     r^F(x)}{2 E_k (\bm{p})} \Theta(E_k (\bm{p})- \mu)
\end{align}
with
\begin{equation}
\label{eq:3dreg_energy}
E_k(\bm{p}) = \sqrt{\bm{p}^2 (1+r^F(x) )^2 + m^2} \ .
\end{equation}
The $\mu$-dependence is solely determined by the Heaviside step
function and since $E_k(\bm{p})>m$ for all momenta the Silver Blaze
property is fulfilled as long as $\mu < m$.

We end this section with a remark: Although a dimensionally reduced 3d
regulator breaks $O(4)$ Euclidean spacetime symmetry only a minor
quantitative impact on the thermodynamics is expected. Moreover, at
very large densities it might be advantageous to waive the Silver
Blaze property in favor of other regulator characteristics. This is
discussed in \cite{Braun:2020bhy} where a symmetric summation of the
fluctuations around the Fermi surface with a Silver Blaze--violating
regulator has been found to improve the findings for BCS-like
theories.

\section{Application: Quark-Meson Model}
\label{sec:qm_model}

As an explicit application of the above considerations we employ a
chiral two-flavor quark-meson model for the effective action with both
bosonic fields $\sigma$, $\bm{\pi}$ as well as fermionic fields
$\psi$. This model is widely regarded as an effective low-energy
truncation to QCD, cf.~Refs. \cite{Jungnickel1995, Berges:1998sd,
  Schaefer:2006ds, Schaefer:2006sr, Braun:2014ata, Zhang:2017icm,
  Rennecke:2016tkm, Fu:2022gou}. Its action
\begin{align}
\label{eq:16}
  \begin{split}
\Gamma_k[\phi,\bar{\psi},\psi] &= \int d^4 x \left[ \frac{Z_{k,\phi}}{2}
  (\partial_\mu \phi)^2 + U_k(\phi^2) - c \sigma \right.\\ 
&\hspace{0.5cm} + \bar{\psi} \left(Z_{k,\psi} \slashed{\partial} +
  \frac{g_k}{2} \left(\sigma + \mathrm{i}\gamma_5 \bm{\tau} \cdot
    \bm{\pi}\right) \right)\psi  \bigg]
\end{split}
\end{align}
includes a field- and scale-dependent effective chiral potential
$U_k (\phi^2)$ for the meson fields. The fields can be combined in the
a $O(4)$-symmetric four-vector $\phi =(\sigma, \bm{\pi})$ such that
the chirally symmetric potential depends only on the radial length
squared $\phi^2$.  Spontaneous breaking of chiral symmetry occurs for
a global minimum at non-zero field value, $\vev{\phi^2} \neq 0$. One
generally chooses for the ground state
$\vev{\phi} = (\sigma_0, \bm{0})$ wherein the radial mode
$\sigma_0 \sim \vev{\bar{\psi}\psi}$ is related to the order parameter for
chiral symmetry breaking, the chiral condensate.

The (constituent) quark field $\psi$ carries $N_c=3$ color degrees of
freedom and interacts with the four (pseudo)scalar mesons, the
$\sigma$-meson and the three pions $\bm{\pi}$, through a color-blind
but running Yukawa coupling $g_k$. The interaction is nondiagonal in
flavor space, signified by the Pauli matrices $\bm{\tau}$, and ensures
isospin and chiral $SU(2)_V \times SU(2)_A$ symmetry.

The constant and scale-independent explicit chiral symmetry
breaking term $-c\sigma$ in \eqref{eq:16} incorporates the effects of finite
current quark masses and yields finite pion
masses. Throughout the work this parameter is fixed to
$c=(120.73 \, \mathrm{MeV})^3$ yielding $m_{\pi} = 138$ MeV.

The scale-dependent but field-independent bosonic and fermionic
wavefunction renormalizations $Z_{k, \phi}$ and $Z_{k, \psi}$ as well
as the Yukawa coupling $g_k$ are needed to determine an approximate
effective potential at the chiral symmetry breaking scale
$k_{\chi}$. In the subsequent solution of the full flow in local
potential approximation (LPA), which corresponds to a leading-order
derivative expansion, they are set constant but the running of the
full effective potential $U_k(\phi^2)$ is taken into account. Details
on the solution procedure are laid out in
Sec. \ref{sec:parameter-fixing}.

\subsection{Regulator Choices}
\label{sec:regulator-choices}

As outlined in Sec.~\ref{sec:regulator-functions}, possible regulator
artifacts might be more apparent at finite chemical potential. In
order to test the impact of the regulator shape function especially on
the back-bending property in the low-temperature phase diagram, three
different regulator functions are considered in the following:
\begin{align}
\label{eq:4dstepreg}
\text{(I)}&&R_k^{\mathrm{mass, 4d}}(p) = k^2 \, \Theta(k_\phi^2 - p^2) \\
\label{eq:3dstepreg}
\text{(II)}&&R_k^{\mathrm{mass, 3d}} (\bm{p}) = k^2 \, \Theta(k_\phi^2-\bm{p}^2) \\
\label{eq:3dflatreg}
\text{(III)}&&R_k^{\mathrm{flat, 3d}}(\bm{p}) = (k^2-\bm{p}^2) \, \Theta(k^2 -
                       \bm{p}^2) 
\end{align}

The regulators in
  Eqs.~(\ref{eq:4dstepreg})-(\ref{eq:3dflatreg}) are bosonic
  regulators. To retain chiral symmetry, the fermionic analogues are
  chosen as outlined in
  Eqs.~(\ref{eq:fermi_regulator})-(\ref{eq:fermi_flat_regulator}). 

\paragraph*{(I)} 

The first regulator is closely related to the Callan-Symanzik
regulator $R_k^\mathrm{CS}(p^2) = k^2$ and will be referred to as
\textit{mass-like} regulator\footnote{Strictly speaking, only
    the bosonic version is directly related to the Callan-Symanzik
    regulator. Due to chiral symmetry, the fermionic regulator has
    an additional non-trivial Dirac structure.}. Due to the momentum-independent
mass-like factor $k^2$ in front of the $\Theta$-function, high momenta
are never fully integrated out and don't decouple for Callan-Symanzik
type flows. Thus, in a strict sense they are not Wilsonian flows,
i.e., the notion of integrating out fluctuations in momentum shells
does not apply. This results in relatively poor performance in
critical exponents computations \cite{Litim:2002cf}. However, since
the momentum dependence of the effective loop propagator
$(\Gamma_k^{(2)} + R_k)^{-1}$ is not modified by such a term, this
regulator represents a reasonable testing ground for the present study
and makes it a natural choice for a 4d regulator since many issues as
discussed in Sec.~\ref{sec:silver_blaze} can be circumvented. The
problem of the missing UV regularization in the Callan-Symanzik
regulator is solved by introducing a step function that suppresses the
four-momenta larger than the compositeness scale $k_\phi$ and serves
as an initial UV scale for the theory. Phenomenologically speaking, at
the compositeness scale the mesonic bound states of quark bilinears
form; see Ref. \cite{Jungnickel:1997ke, Berges:1998sd} for early
applications of this regulator function. Thus, $k_\phi$ presents a
hard physical cutoff scale. Note that in a recent work
\cite{Braun:2022mgx} a novel flowing renormalization procedure was
introduced which allows to cancel the explicit $k_\phi$
dependence. However, the momentum argument of the step function in
\eqref{eq:4dstepreg} turns problematic since an analytic continuation
to complex frequencies for finite chemical potentials is not possible.
Similar as in Ref. \cite{Berges:1998sd} we proceed by temporarily
replacing the $\Theta$-function with a smeared-out version
$\Theta_\epsilon$, assuming that such a continuation then exists. All
Matsubara sums can be solved analytically and the flow for the
potential splits into a vacuum and thermodynamic contribution
\begin{equation}
  \label{eq:6}
  \partial_t U_k(\sigma^2,T,\mu) = \partial_t
  U_k^\mathrm{vac}(\sigma^2) + \partial_t
  U_k^\mathrm{th}(\sigma^2,T,\mu) \ . 
\end{equation}
The vacuum contribution reads for $\nu = 4 N_c N_f$
\begin{align}
	\label{eq:7}
	\begin{split}
		\partial_t U_k^\mathrm{vac} = k^2 \int_p \Theta(k_\phi^2-p^2)
		&\Bigg( \frac{1}{p_0^2+E_\sigma^2}  +
                  \frac{3}{p_0^2+E_\pi^2}  \\ 
		& \hspace{1.5cm} \ -
		\frac{\nu}{p_0^2+E_\psi^2} \Bigg)
	\end{split}
\end{align}
with the  quasi-particle energies 
$E_i(\bm{p}) = \sqrt{\bm{p}^2+k^2+m_i^2}$ for the fields $i
\in\{\sigma,\pi,\psi\}$. 
The corresponding masses are $m_\sigma^2 = 2U_k' + 4\sigma^2 U_k''$,
$m_\pi^2 = 2 U_k'$ and $m_\psi^2 = (g\sigma/2)^2$, wherein derivatives
with respect to $\sigma^2$ are denoted by a prime, i.e.,
$U_k' \equiv d U_k/d \sigma^2$.
An analytical integration of \eqref{eq:7} is possible and straightforward.

The thermal part is given by
\begin{align}
	\begin{split}
		\label{eq:4d_thermal_flow}
		\partial_t U_k^\mathrm{th} &= k^2 \int_{\bm{p}}
		\left(\frac{n_B(E_\sigma,T)}{E_\sigma} +
		\frac{3n_B(E_\pi,T)}{E_\pi}\right. \\ 
		&\hspace{0.5cm}\left.+ \frac{\nu \,
			[n_F(E_\psi,T,\mu)+n_F(E_\psi,T,-\mu)]}{2E_\psi}\right) \ ,
	\end{split}
\end{align}
where $n_B$ and $n_F$ denote the standard Bose and Fermi distributions
\begin{align}
\label{eq:8a}
\begin{split}
	n_B(E,T) &= \frac{1}{\mathrm{e}^{E/T}-1} \ , \\
	\quad n_F(E,T,\mu) &= \frac{1}{\mathrm{e}^{(E-\mu)/T}+1} \ .
\end{split}
\end{align}

Since only the vacuum flow requires a UV regularization the
compositeness scale is not necessary in the UV finite thermal flow
contribution. This means that the discontinuity of the step function
can formally be sent to infinity, $k_\phi \rightarrow \infty$,
and any contributions from additional poles of the smeared
$\Theta_\epsilon$ function can be safely ignored.

\paragraph*{(II)}

For the three-dimensional version of the mass-like regulator
$R_k^{\mathrm{mass, 3d}}$, \eqref{eq:3dstepreg}, the vacuum part of
the flow does not exhibit the complete $O(4)$ symmetry anymore:
\begin{equation}
\label{eq:3d_vacuum_flow}
\partial_t U_k^\mathrm{vac} = \frac{k^2}{2} \int_{\bm{p}}
\Theta(k_\phi^2-\bm{p}^2) \left(\frac{1}{E_\sigma} + \frac{3}{E_\pi} -
  \frac{\nu}{E_\psi} \right) \ . 
\end{equation}
The thermal flow contribution is identical to the previous 4d version,
\eqref{eq:4d_thermal_flow}, with the exception that we do not take the
limit $k_\phi \rightarrow \infty$ because the $\Theta$-function only
acts on spatial momenta and does not introduce any additional poles.

It should be noted that both mass-like regulators are not optimized
according to any optimization criteria as discussed in
App.~\ref{app:optimization}.

\paragraph*{(III)}

The third regulator is the often-used flat regulator in three
dimensions, also known as Litim regulator. It removes all
spatial-momentum dependence from the loop propagator, such that the
quasi-particle energies for the field $i$ turn into
$E_i = \sqrt{k^2+m_i^2}$.

In contrast to Callan-Symanzik type flows, at a given scale $k$ all
fluctations with (spatial) momenta larger than $k$ are completely
integrated out. The vacuum and thermal contributions take the simple
forms
\begin{equation}
\label{eq:vacuum_flow_litim3d}
\partial_t U_k^\mathrm{vac} = \frac{k^5}{12 \pi^2}
\left(\frac{1}{E_\sigma} + \frac{3}{E_\pi} - \frac{\nu}{E_\psi} \right)
\ , 
\end{equation}
and
\begin{align}
\begin{split}
\label{eq:thermal_flow_lititm3d}
\partial_t U_k^\mathrm{th} &= \frac{k^5}{6 \pi^2}
\left(\frac{n_B(E_\sigma,T)}{E_\sigma} + \frac{3 n_B(E_\pi,T)}{E_\pi}
\right. \\ 
&\hspace{0.5cm} \left.+ \frac{\nu \, (n_F(E_q,T,\mu) + n_F(E_q,T,-\mu))}{2 E_\psi} \right) \ .
\end{split}
\end{align}

\subsection{Parameter Fixing}
\label{sec:parameter-fixing}

For the explicit numerical solution of the flow equations an initial
action needs to be fixed in the UV. Usually, in LPA the potential
$U_{k=\Lambda}$ is parameterized by some couplings for a given UV
cutoff to reproduce physical observables in the infrared.
Unfortunately, for the mass-like regulators this procedure could not
be applied. A fixing of the potential up to some quartic couplings
with a satisfactory chiral symmetry breaking scenario in the infrared
was not possible since the needed numerical parameter space was not
accessible. The pole structure of the threshold
functions in the corresponding flow equations impedes the numerical
handling in particular for the vacuum flow and close to the origin of
the radial $\sigma$-field. In App.~\ref{app:pole_proximity} more
details and consequences of the pole proximity for different
regulators are given.

However, we circumvent this issue and fix the initial action as
follows:
One feature of the quark-meson model truncation is the presence of a
sort of an approximate partial IR fixed point in the symmetric regime
above the chiral symmetry breaking scale $k_{\chi}$. All trajectories
of the running couplings with initial values fixed at scales larger
than $k_{\chi}$ show a similar convergence behaviour.
The partial fixed point behavior is inherited for heavy mesons from
effective four-quark interactions generated in QCD \cite{Gies:2002hq,
  Braun:2014ata, Fu:2019hdw} and can be used to fix the effective
potential at $k_\chi$, see also \cite{Berges:1997eu,
  Berges:1998sd}. Due to the fixed-point behavior many infrared
parameters of the action will be almost independent of their initial
values since the system eventually loses its memory of the initial
values fixed at the larger compositeness scale $k_{\phi} >
k_{\chi}$. In addition, for large enough Yukawa couplings only a few
relevant parameters need to be determined from QCD or alternatively
from phenomenology. The physical IR fixed point at $k\to 0$ can be
estimated for scales in the symmetric regime $k_{\chi} < k < k_{\phi}$
with the flow equations for both wavefunction renormalizations
$Z_{k, \phi}$ and $Z_{k, \psi}$, the Yukawa coupling $g_k$, and the
effective potential $U_k(\phi^2)$.  From the Landau pole of the
renormalized Yukawa coupling in the vicinity of the compositeness
scale $k_{\phi}$ the mesons have large renormalized masses and the
flows are dominated by the quarks following from the condition
$Z_{k_{\phi}, \phi} \ll 1$.

It is therefore reasonable to consider only the purely fermionic
contributions to the flows.  The Yukawa coupling feeds back to the
flow of the meson wavefunction renormalization whereas the fermionic
anomalous dimension vanishes. The corresponding solution for the
dimensionless effective potential
\begin{equation}
\label{eq:dimensionless_ren_potential}
u_k(\tilde{\rho}) := \frac{U_k(\phi^2)}{k^4}
\end{equation}
as a function of the dimensionless
renormalized chiral invariant
\begin{equation}
\label{eq:dimensionless_chiral_invariant}
\tilde{\rho} := Z_{\phi,k} \frac{\phi^2}{2k^2}
\end{equation}
exhibits a partial infrared fixed point.

Assuming a power expansion of the potential around the origin
\begin{equation}
\label{eq:power_expansion_uk}
u_k(\tilde{\rho}) = \sum_{n=0}^\infty \frac{1}{n!}
u_k^{(n)}(0) \, \tilde{\rho}^n \ ,
\end{equation}
infrared-attractive solutions for the scaled coefficients 
\begin{equation}
\frac{u_k^{(2)}(0)}{\bar{g}_k^2} \quad \mathrm{and} \quad
\frac{u_k^{(n)}(0)}{\bar{g}_k^{2n}} \quad \mathrm{for} \ n>2\   
\end{equation}
can be found, where $\bar{g}_k=Z_{k,\psi}^{-1} Z_{k,\phi}^{-1/2} g_k$
is the renormalized Yukawa coupling, cf. \cite{Berges:1997eu} for a
similar analysis. Further technical details of the fixed-point
solution are moved to App. \ref{app:approximate_flows}.  At $k_\chi$
the coefficient $u_{k_{\chi}}^{(1)}(0)$ vanishes and the approximation
of large renormalized meson masses breaks down. The other coefficients
$u_{k=k_\chi}^{(n)}(0)$ for $n \leq 4$ are set to their infrared
fixed-point values and all higher orders $n>4$ are neglected. The
$n=0$ coefficient is just a constant and can be ignored. From $k_\chi$
downwards, the Yukawa coupling starts to freeze out. Hence, from this
scale on the full flow is solved in LPA for a fixed Yukawa coupling
$g = \bar{g}_{k=k_\chi} = 6.5$ and vanishing anomalous dimensions. In
principle, the only free parameter left is the symmetry breaking scale
$k_\chi$. It can be fixed to yield the physical pion decay constant,
$\sigma_\mathrm{min} \approx f_\pi = 92.4 \, \mathrm{MeV}$ in the
infrared.

For the mass-like regulators, Eqs.~(\ref{eq:4dstepreg}) and
(\ref{eq:3dstepreg}), $k_\phi$ is an additional free parameter. The
correct combination of the two parameters is not so clear since, for
example, the sigma mass seems to only weakly depend on it.

The chosen parameter sets that seem to lie in a physically acceptable
region are tabulated in Tab. \ref{tab:parameters}. The
$n=2$ coefficient is regulator independent and is fixed to one at
$k_{\chi}$. Note that the direct computation of the vacuum flow is not
possible for the mass-like regulators due to the mentioned pole
structure in the flow equation (see discussion in
App. \ref{app:pole_proximity}). A numerical treatment, however, is
possible at small temperatures and chemical potentials around the
(pseudo)critical value $\mu_c$. By extrapolating into the Silver Blaze
region we thus can infer approximate vacuum solutions. For the
mass-like regulators we could not determine a precise vacuum sigma
meson mass in this manner due to a strong $\mu$ dependence around the
critical $\mu_c$. For both regulators we nevertheless expect mass
values around $m_\sigma = 510$ MeV similar as the ones found for the
flat regulator within an error of about 40 MeV.

Note that the assumption of a $(T,\mu)$-independent initial UV action
only holds for sufficiently low external parameters much smaller than
$k_{\chi}$, i.e. $T \ll k_\chi$ as well as $\mu <
k_\chi$ \cite{Braun:2018svj}.

Finally, we would like to emphasize that the sufficient memory loss of
the initial values, i.e., the close proximity to the fixed point, is
not yet necessarily satisfied at the chiral scale $k_{\chi}$ in
contrast to the statements in \cite{Berges:1997eu}. Therein it is
indicated that the IR fixed point is already established at the chiral
scale. However, for a ratio $k_\chi/k_\phi \approx 1/2$, which
corresponds to an RG time of just $t\approx - \ln (2)$, the solutions
show that the $t\rightarrow -\infty$ fixed point is not nearly reached
at this point.

Furthermore, the explicit $k_\phi$-dependence of the mass-like
regulators leads to a modification of these solutions. Approximate IR
fixed points still appear when $k_\phi/k \rightarrow \infty$, but the
convergence to these values might be even slower. Nevertheless, we
regard the explained procedure as a good choice for a testing ground,
allowing for the comparison of different regulator schemes within a
common setup and with only a few free parameters.

\begin{table}
\centering
\begin{tabular}{| c | c | c | c | c |}
\hline
 & $k_\chi$ [MeV] & $k_\phi$ [MeV] & $u_{k_\chi}^{(3)}/\bar{g}_{k_\chi}^6$ & $u_{k_\chi}^{(4)}/\bar{g}_{k_\chi}^8$\\
 \hline
\, mass, 4d \, & 480 & 690 & \ -0.00950 \ & \ 0.00475 \ \\
\, mass, 3d \, & 390 & 610 & \ -0.00950 \ & \ 0.00475 \ \\
\, flat, 3d \, & 580 & $-$ & \ -0.02375 \ & \ 0.02078 \ \\
\hline
\end{tabular}
\caption{Chiral symmetry breaking scale $k_{\chi}$ and compositeness
 scale $k_{\phi}$ as well as starting parameters of the effective
  potential for the three different regulators. Evaluation of the
  $u_k^{(n)}$ at $\tilde{\rho}=0$ is implied. Note that the infrared
  solutions for the mass-like regulators degenerate and the $n=2$
  infrared attractive point is regulator independent resulting in
  $u_{k_\chi}^{(2)}/\bar{g}_{k_\chi}^2 \equiv 1$.
  \label{tab:parameters} }
\end{table}

\section{Numerical Results}
\label{sec:numerical_results}

\begin{figure}
  \centering
  \includegraphics[width=\onefig]{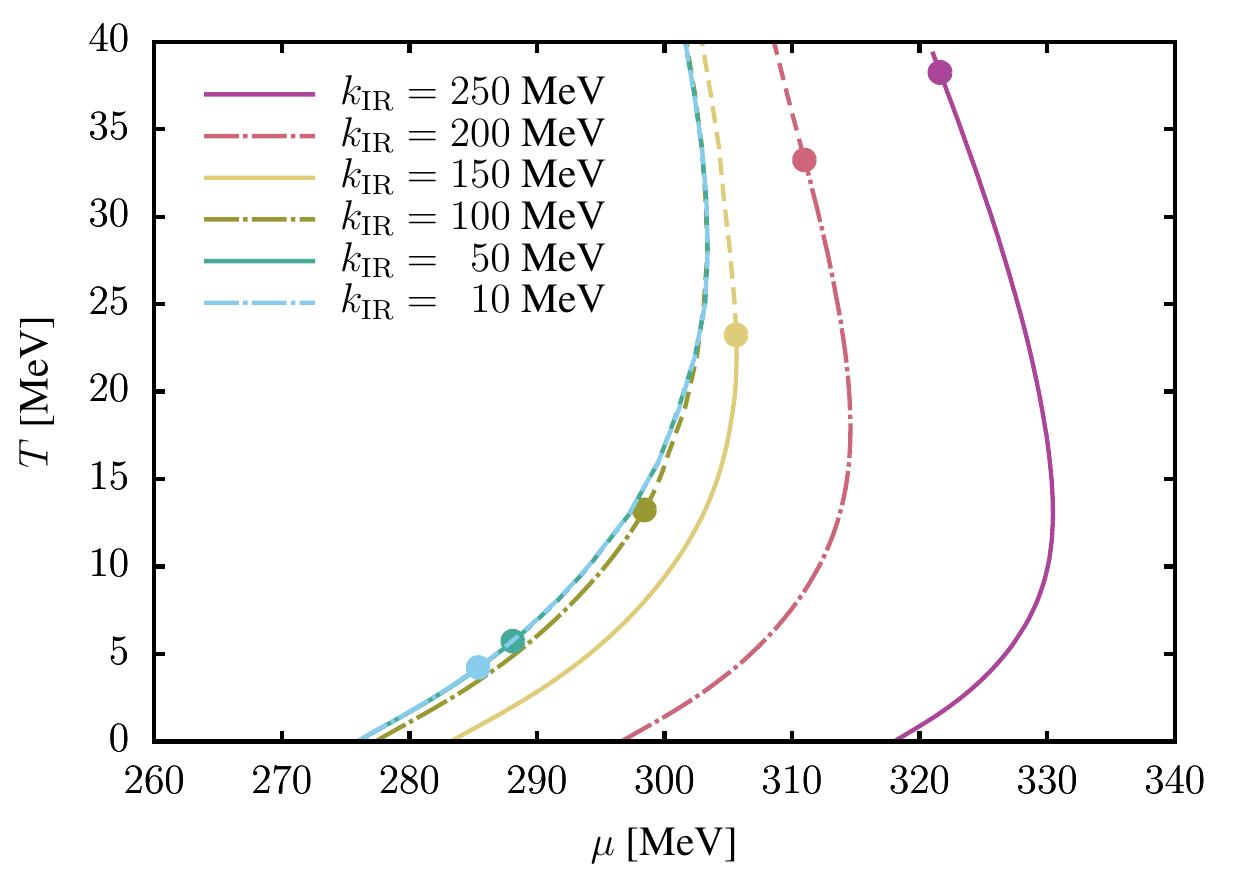} 
  \caption{\label{fig:litim3d_kir} Infrared cutoff dependence of the
    chiral phase transition with the 3d flat regulator. Critical
    endpoints are marked by dots, solid and dash-dotted lines denote first-order and
    dashed lines crossover transitions.}
\end{figure}

In this section, we focus on the phase boundary of the chiral phase
transition at high densities. One of the first questions concerns the
impact of a finite infrared cutoff $k_\mathrm{IR} > 0$ in the flow. In
Fig. \ref{fig:litim3d_kir} the infrared scale dependence of the chiral
phase diagram for the 3d flat regulator is demonstrated. A familiar
back-bending of the transition line as well as a movement of a
critical endpoint towards lower temperatures is found with the
consequence that the critical chemical potential moves to lower
values. The back-bending occurs even at large infrared scales and is
not related to the location of the critical endpoint.  The flow in the
infrared is dominated by the lightest degrees of freedom, the pions,
which tend to restore the chiral symmetry. As a consequence, for
scales below the pion mass, $k_\mathrm{IR} \lesssim 100$ MeV, the
phase boundary consolidates.  The movement of the critical endpoint
can be traced back to the value of the sigma meson curvature mass in
the infrared \cite{Schaefer:2008hk, Schaefer:2006ds} which does not
freeze due to the running of the second derivative of the potential
$U''_k$.

Hence, the back-bending is a generic feature of the employed
truncation and regulator and not of the particular choice of the UV
potential or IR cutoff. In the following, we will always use a fixed
$k_\mathrm{IR} = 50$ MeV.

\begin{figure}
  \centering
  \includegraphics[width=\onefig]{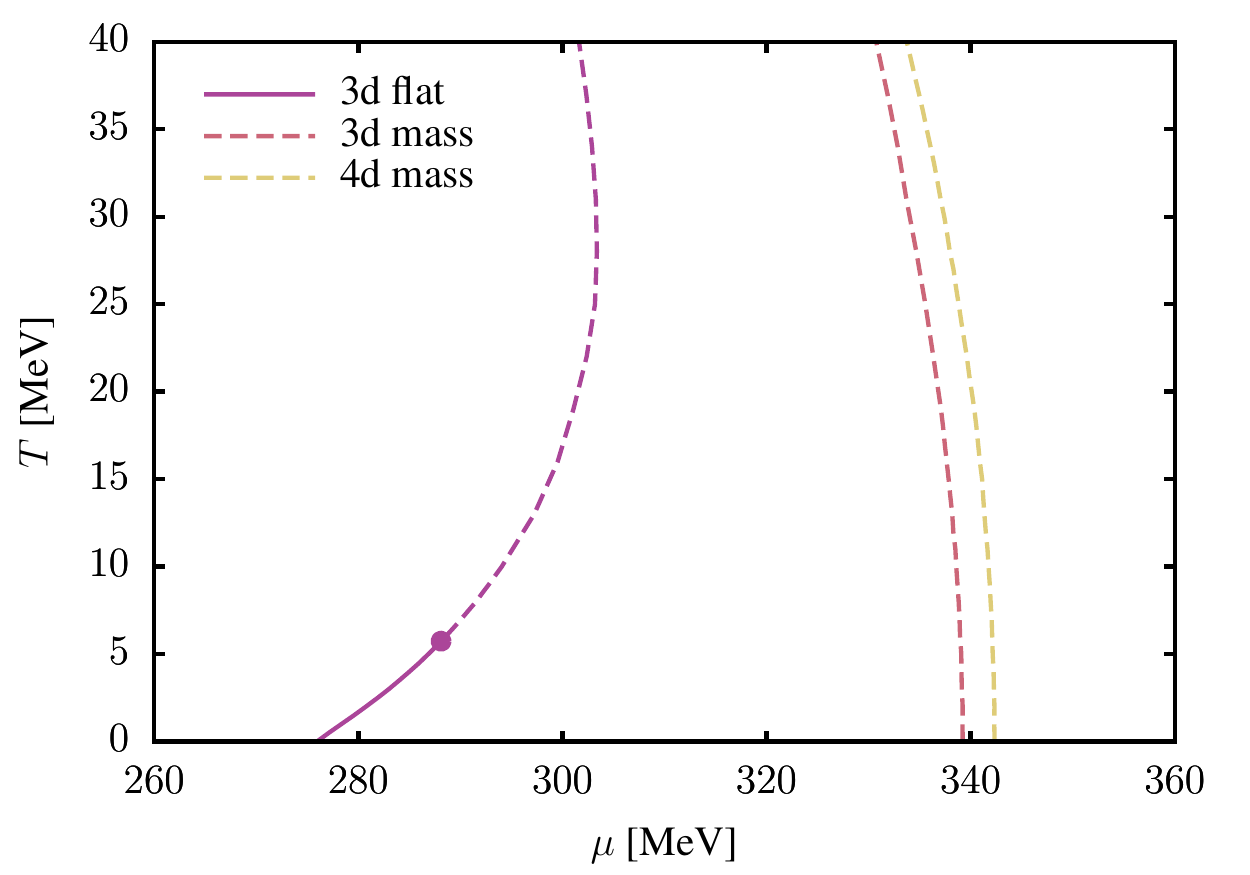} 
  \caption{\label{fig:phasediag_comparison} Regulator scheme dependence of
    the chiral transition for three different regulators. Line styles 
    similar to Fig.~\ref{fig:litim3d_kir}.}
\end{figure}

\begin{figure*}
  \centering
  \subfigure[\label{fig:litim3d_entropy} 3d flat regulator.]{\includegraphics[width =
    \twofigs]{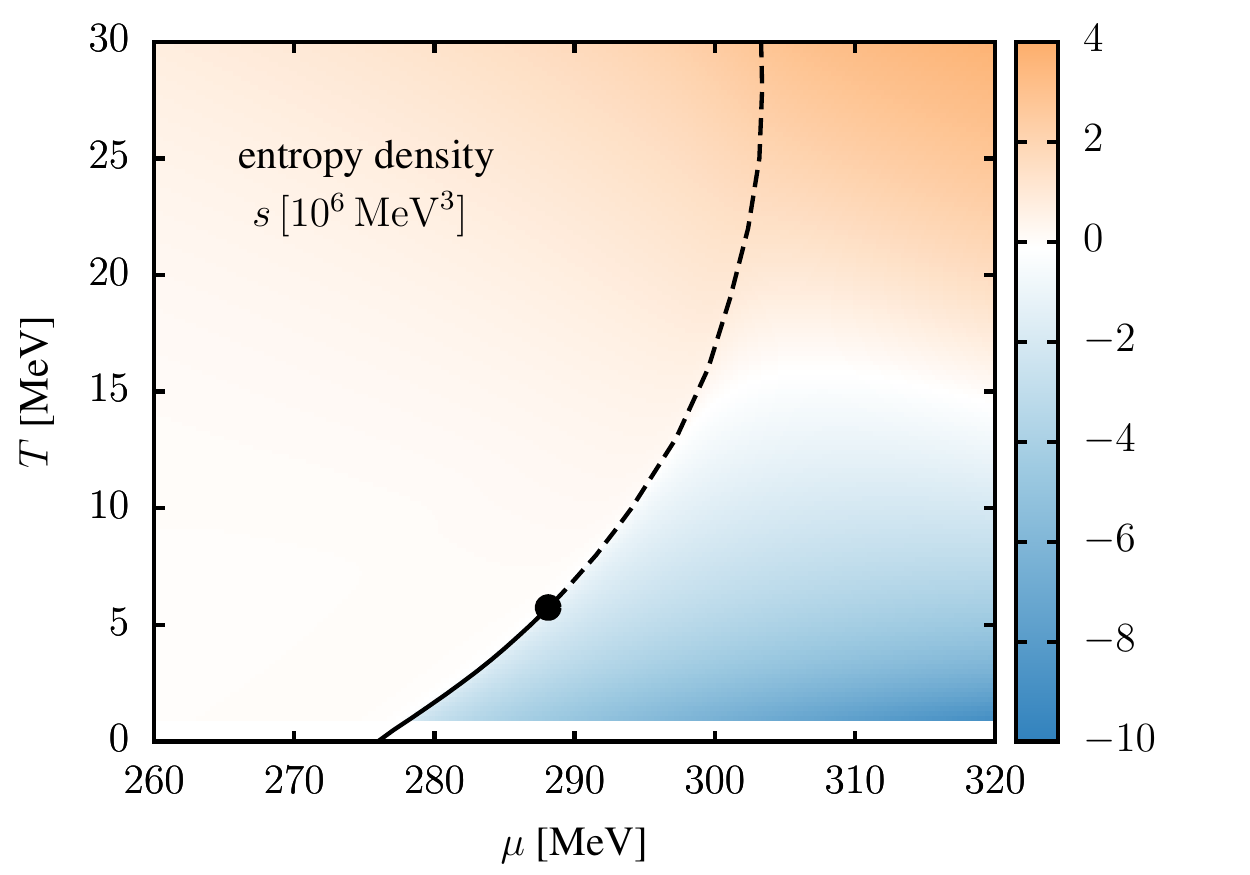}} 
\subfigure[\label{fig:power3d_entropy} 3d mass-like regulator.]{\includegraphics[width =
    \twofigs]{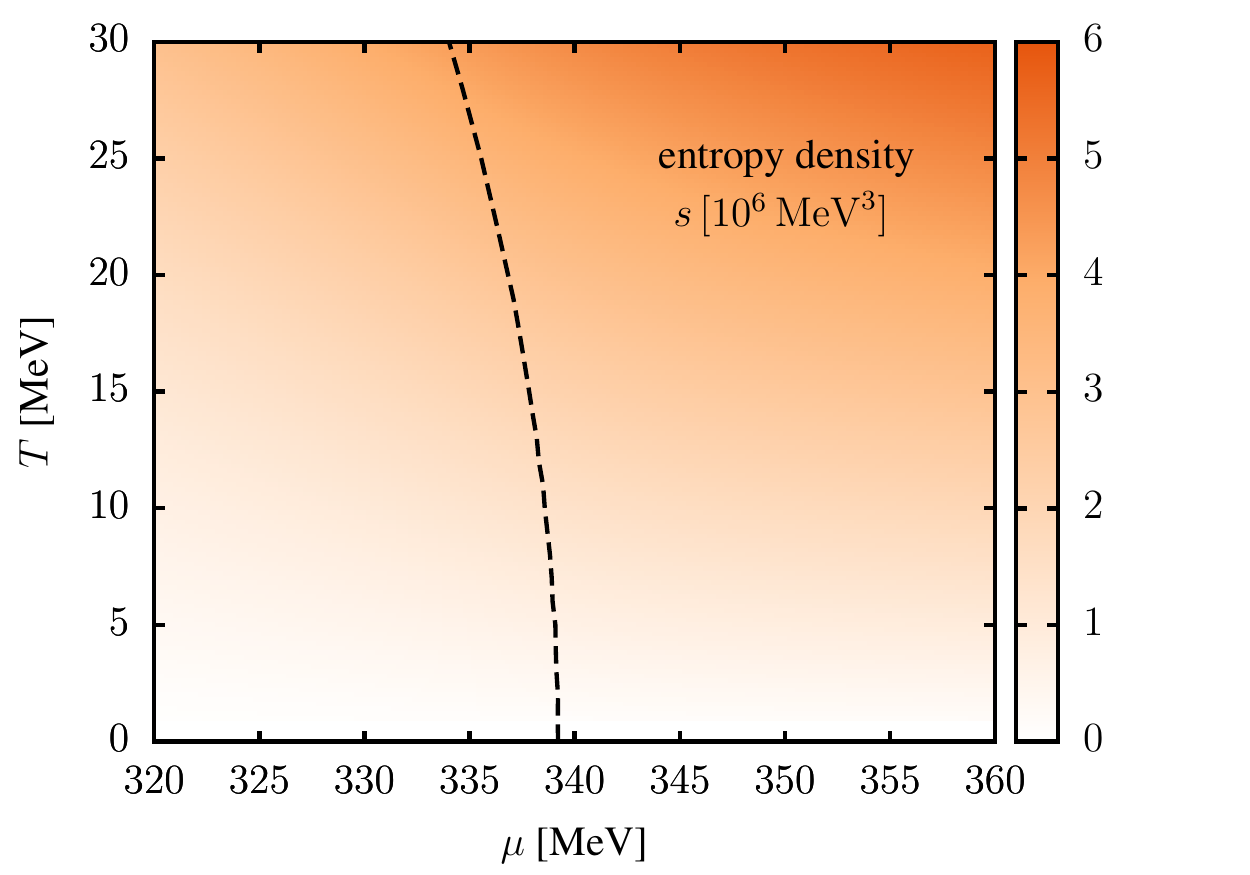}} 
  \caption{\label{fig:entropy_total} Entropy density for the 3d flat
    (left panel) and 3d mass-like (right panel) regulators close to
    the zero-temperature chiral phase boundary. The lower blue-shaded region denotes
    a negative entropy density. }
\end{figure*}

In Fig. \ref{fig:phasediag_comparison} the regulator scheme dependence
of the chiral phase boundary at low temperatures is presented. While
the phase boundary obtained with the common 3d flat regulator
exhibits a back-bending it vanishes for mass-like regulators. For
both the 3d and 4d versions, the transition line hits the $\mu$-axis
perpendicularly. This is already a strong hint that the back-bending
in LPA is related to the momentum structure of the employed
regulator. For both mass-like regulators the chiral transition is a
smooth crossover and the critical endpoint which is still present at
$T \approx 6$ MeV for the flat regulator is gone (basically pushed
below the $\mu$-axis); this behavior should not be over-interpreted
since the location of the critical endpoint strongly depends on
parameter choices \cite{Schaefer:2008hk}.  Both crossovers are closely
aligned at $T=0$: $\mu^\mathrm{(cross, 3d)} \approx 339$ MeV and
$\mu^\mathrm{(cross, 4d)} \approx 342$ MeV, respectively. The
difference of about 3 MeV is not of significance since no particular
fine-tuning of the starting parameters for the chiral condensates has
been taken into account. Since there are no qualitative differences
between the two crossovers it seems that dimensionally reduced
regulators are an appropriate choice for the analysis of finite-$\mu$
thermodynamics.

In the chiral limit when the explicit symmetry breaking term $c$ in
the action vanishes, a first-order phase transition for both mass-like
regulators is found. The critical chemical potential is smaller but
always above the vacuum quark mass in this case. Interestingly, this
is in accordance with a similar work \cite{Berges:1998sd} where a
different exponential regulator was employed for the bosonic
fluctuations. Contrarily, with the 3d flat regulator the phase
transition at vanishing temperature is always of first-order
regardless of the explicit symmetry breaking and the critical chemical
potential is smaller than the vacuum quark mass.

\begin{figure}
  \centering
  \includegraphics[width=\onefig]{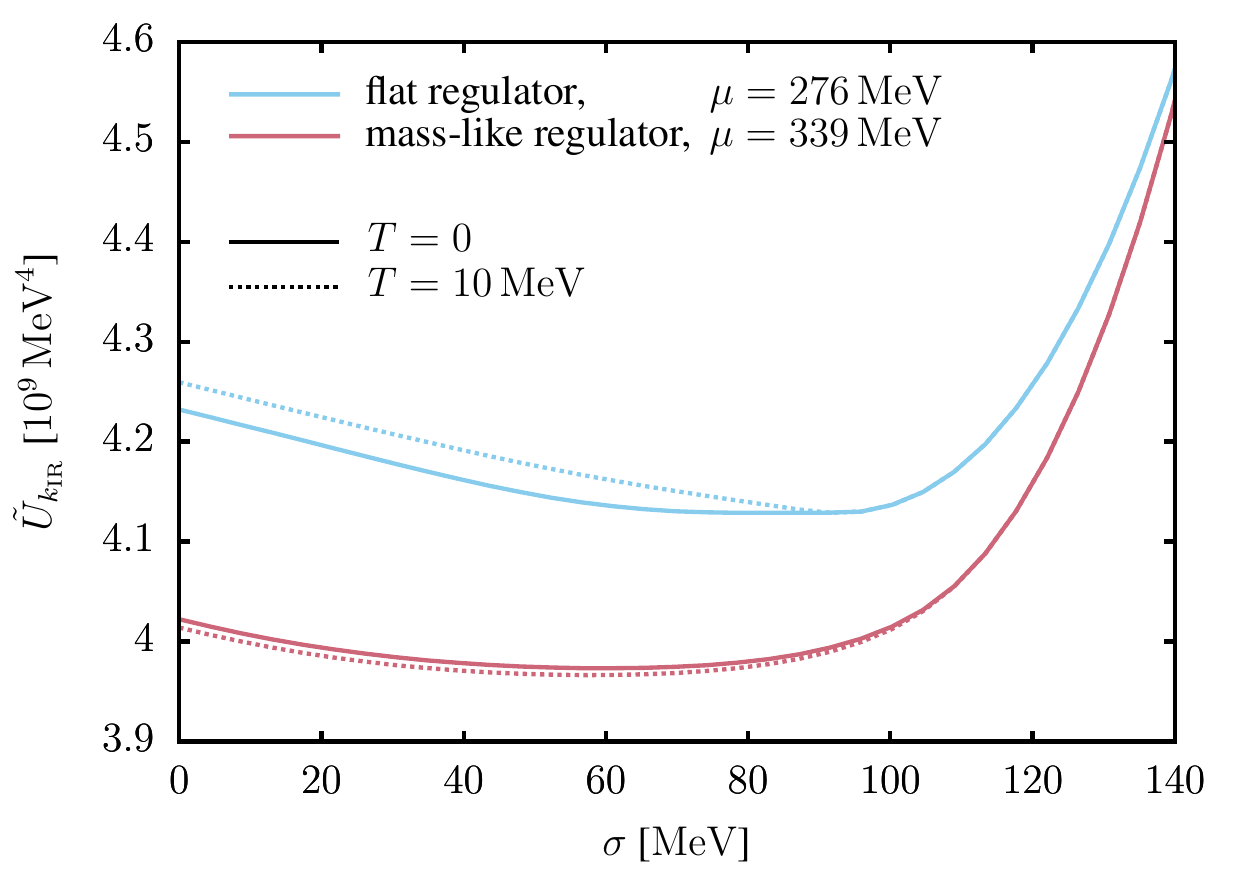} 
  \caption{\label{fig:potential_total} 3d flat and 3d mass-like
    effective potentials with explicit symmetry breaking at
    $k_\mathrm{IR} = 50 \, \mathrm{MeV}$ for $T=0$ (solid) and
    $T=10 \, \mathrm{MeV}$ (dotted) as a function of the radial
    $\sigma$-mode. The chemical potential is fixed to the respective
    zero-temperature transitions:
    $\mu_c^\mathrm{(1st-order)} \approx 276$ MeV for the flat and
    $\mu^\mathrm{(cross)} \approx 339$ MeV for the mass-like
    regulator. The mass-like regulator potential has been shifted by a
    constant for comparison. }
\end{figure}

Characteristic for the back-bending is the occurrence of a negative entropy
density $s$ beyond the chiral transition line at small
temperatures. As already discussed in \cite{Tripolt2017} the
back-bending is in agreement with the Clausius-Clapeyron relation
\begin{equation}
\label{eq:2}
\frac{dT_c}{d\mu_c} = - \frac{\Delta n}{\Delta s}\ ,
\end{equation}
since a positive particle density difference $\Delta n$ and a negative
entropy density difference $\Delta s$ result in a finite positive
slope of the (first-order) transition line on the $\mu$-axis. A
comparison of the thermodynamics is given in
Fig.~\ref{fig:entropy_total} where the entropy density in the
low-temperature region of the phase diagram for the two 3d
regulators is shown. The left panel shows the expected negative
entropy density (blue shaded region) for the flat regulator
\cite{Tripolt2017}.  With the mass-like regulator (right panel) the
negative entropy density vanishes along with the back-bending.

\begin{figure*}
  \centering
  \subfigure[\label{fig:threshold_litim3d} 3d flat regulator.]{\includegraphics[width =
    \twofigs]{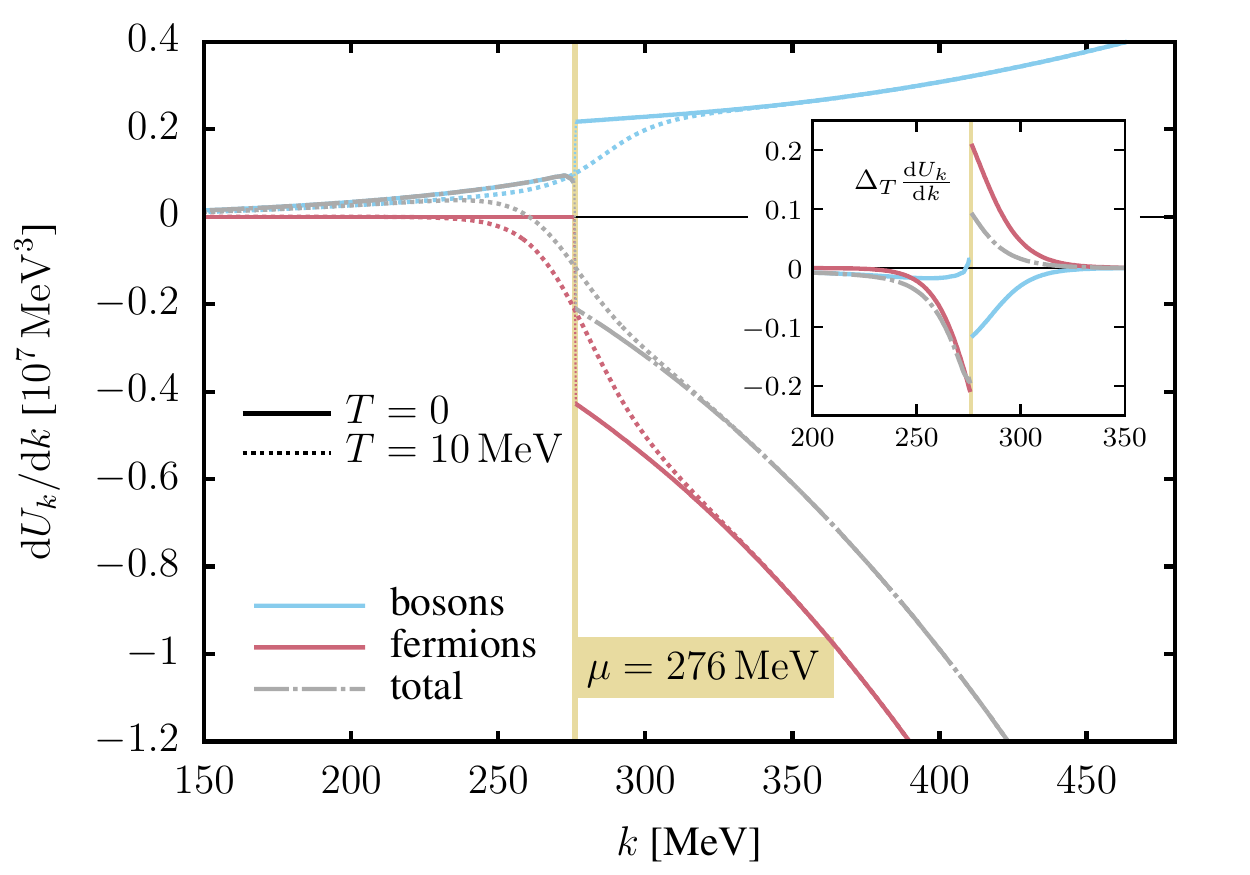}} 
\subfigure[\label{fig:threshold_power3d} 3d mass-like regulator.]{\includegraphics[width =
    \twofigs]{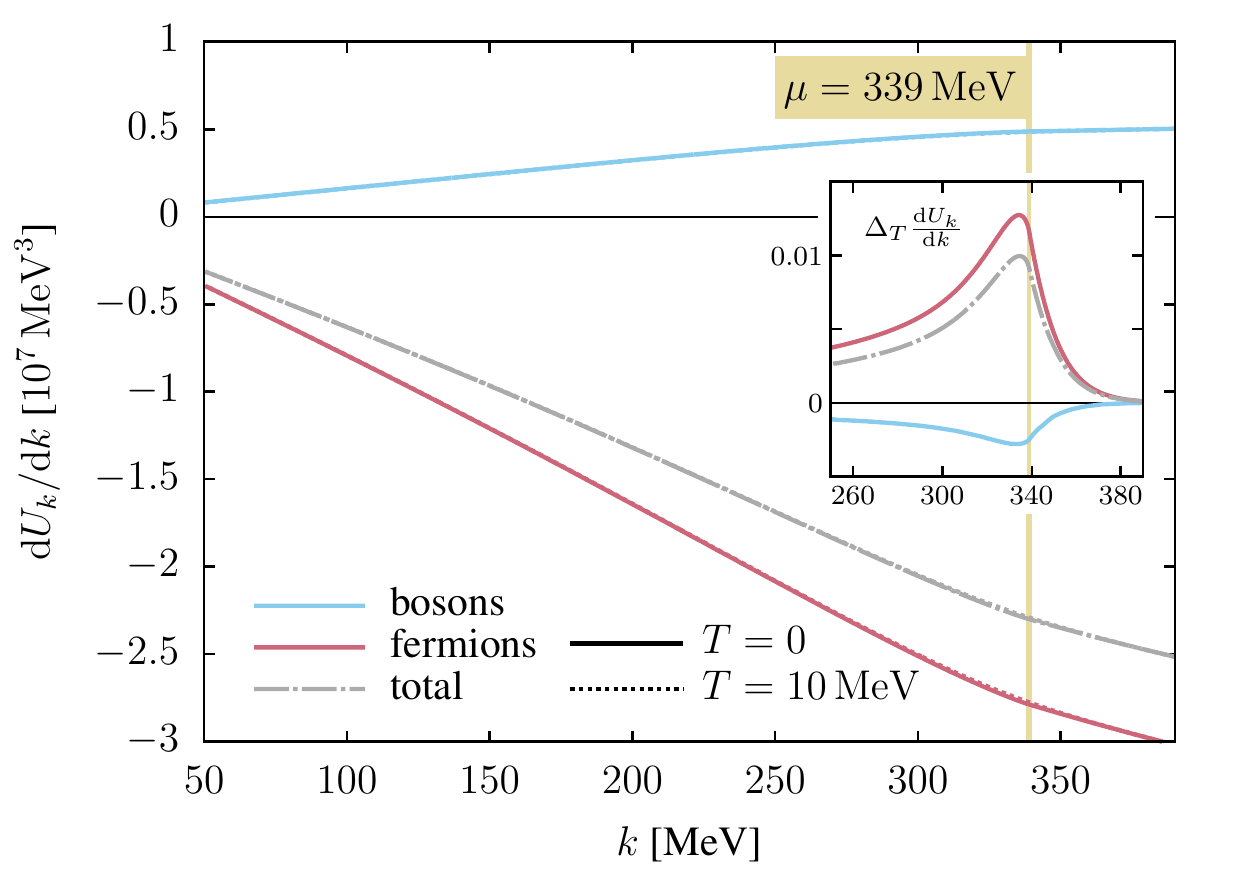}} 
  \caption{\label{fig:threshold_total}
   Contributions to the potential flow 
   for $T=0$ (solid and dash-dotted) and $T=10$ MeV (dotted) at
   $\sigma = 0$ as a function of the RG scale $k$. Left: 3d flat
   regulator; right: 3d mass-like regulator. The inlay shows the
   respective $T=0$ flows subtracted from the $T=10$ MeV flows. Chemical
   potentials (vertical lines) are chosen as in
   Fig. \ref{fig:potential_total}. }
\end{figure*}

The notably different back-bending behavior between the 3d flat and
the 3d mass-like regulators is also visible in the fully evolved
effective potential. In Fig.~\ref{fig:potential_total} both effective
potentials
$\tilde{U}_{k_\mathrm{IR}} \equiv U_{k_\mathrm{IR}} - c \sigma$, i.e.,
with an explicit symmetry breaking term and evolved to
$k_\mathrm{IR} = 50$ MeV, are shown as a function of the radial
$\sigma$-mode for temperatures $T=0$ (solid lines) and $T=10$ MeV
(dotted lines). The corresponding chemical potentials are each fixed
close to the transition at vanishing temperature: for the flat
regulator $\mu_c^\mathrm{(1st-order)} \approx 276$ MeV (blue colors)
and for the 3d mass-like regulator $\mu^\mathrm{(cross)} \approx 339$
MeV (red colors) where the latter potential has been shifted by an
irrelevant constant for a better comparison. One sees that the
infrared potential for the mass-like regulator decreases for
increasing temperature while the opposite behavior is found for the 3d
flat regulator where the potential increases with increasing
temperature.  This in turn means that there is stronger spontaneous
chiral symmetry breaking for the flat regulator since the potential
minimum is pushed to larger values for increasing temperature. This
causes the back-bending of the transition line.  For the mass-like
regulator the potential minimum decreases slightly with increasing
temperature such that the chiral transition shifts to smaller chemical
potentials and thus no back-bending occurs.

This different temperature progression of the potentials can be
further traced back to a different flow behavior of the fermion
contributions which contain the explicit $\mu$-dependence.  At small
temperatures and for increasing chemical potential, the thermodynamic
contributions progressively cancel out the vacuum flow, generally
leading to a decoupling behavior of the fermions.

For the 3d mass-like regulator the zero-temperature fermionic flow,
Eqs. (\ref{eq:3d_vacuum_flow}) and (\ref{eq:4d_thermal_flow}),
can be rewritten as
\begin{align}
  \label{eq:9}
\begin{split}
\partial_t U_k^{F,\mathrm{ mass}} = - \frac{\nu k^2}{4 \pi^2} \int_{p_F}^{k_\phi} \mathrm{d}\vert
\bm{p} \vert \, \frac{\bm{p}^2}{E_\psi} 
\end{split}
\end{align}
with the quark energy
$E_\psi = \sqrt{\bm{p}^2 + m_{\text{eff}, \psi}^2}$. Due to the
regulator's momentum independence, the quark energy $E_{\psi}$ looks
like an ideal or free dispersion relation depending on the spatial
momenta with an effective scale-dependent quark mass
$m_{\text{eff}, \psi} = \sqrt{k^2 + m_{\psi}^2}$.  The lower boundary
of the momentum integration defines a Fermi momentum
\begin{equation}
  \label{eq:10}
p_F := \begin{cases}
\sqrt{\mu^2 -  m_{\text{eff}, \psi}^2} \ , & \mu^2 > m_{\text{eff}, \psi}^2 \\
0 \ , & \mathrm{else} \ .
\end{cases}
\end{equation}
For $\mu > m_\psi$ (not $m_{\text{eff}, \psi}$) and $k$ sufficiently
small, the momentum space in the integral is gradually suppressed by
occupied quark states below the Fermi surface and the fermions
decouple from the flow. In other words, the chemical potential serves
as an effective infrared cutoff.

This is in contrast to the flat regulator where the decoupling is not
gradual but occurs at a sharp RG scale $k_F$. For this regulator the
fermionic flow, Eqs. (\ref{eq:vacuum_flow_litim3d}) and
(\ref{eq:thermal_flow_lititm3d}), is a sharp step function at
vanishing temperature
\begin{equation}
\partial_t U_k^{F, \mathrm{ flat}} = -\frac{\nu \, k^5}{12 \pi^2} \,
\Theta(E_\psi - \mu) 
\end{equation}
with energies $E_\psi = \sqrt{k^2 + m_\psi^2}$ which do not depend on
spatial momenta. Here, for $\mu > m_\psi$, the flow discontinuously
jumps to zero at $k_F = \sqrt{\mu^2 - m_\psi^2}$.

A visualization of the flow around the Fermi surface is given in
Fig. \ref{fig:threshold_total}, on the left for the 3d flat regulator
and on the right for the 3d mass-like regulator. In each panel three
different flows evaluated at $\sigma=0$ are shown as a function of the
RG scale. The total flow $\partial_k U_k$ (gray lines) assembles the
bosonic (blue upper lines) and fermionic (red lower lines)
flows. Solid and dash-dotted lines are the flows at zero temperature
and the dashed lines the ones for $T=10$ MeV.  The 3d flat regulator
(left) induces via the $\Theta$-function a discontinuous jump at $T=0$
in all three flows at the critical chemical potential. The fermionic
flow jumps to zero at $k_F = \mu$ since the quark mass $m_{\psi}$
vanishes for $\sigma =0$. Consequently, the total flow changes sign
and is given completely by the bosonic contribution that tries to
restore chiral symmetry. This behavior is smeared out at finite
temperatures and the contribution of the bosons weakens overall. As a
net effect chiral symmetry breaking becomes stronger at finite
temperatures which finally yields the back-bending of the transition
line in the phase diagram.

In the inlays of both panels of \Fig{fig:threshold_total} the
temperature difference $\Delta_T (d U_k/dk)$ of the zero temperature
flow subtracted from the $T=10$ MeV flow is displayed for the same
three potential contributions (same linestyle is used). For the flat
regulator (left inlay) an expected discontinuity arises at the Fermi
surface and the sign of the enclosed area of the fermionic flow
difference and hence total flow contribution changes. Since the areas
of the fermionic flow are almost of the same magnitude around the
critical chemical potential the fermionic flow decouples from the
further evolution towards the infrared. The consequence is that the
net total flow is dominated by the asymmetric bosonic
contributions. Their difference is negative everywhere and pushes the
potential contribution to larger values which finally drives the
chiral symmetry breaking.

This behavior is in contrast to the one with a 3d mass-like regulator
(right panel) where a smooth decoupling takes place around the
corresponding smooth transition. The total flow stays negative and is
almost insensitive to small temperature variations, which in turn
produces the perpendicular evolution of the smooth transition line in
the phase diagram.

Similarly, in the right inlay of \Fig{fig:threshold_total} no sharp
decoupling of the fermionic flow around the Fermi surface is
found. The flow differences peak close to the Fermi scale $k_F$ but
contribute to the total flow for all scales (despite the different
order of magnitudes in the inlays). Since the difference in the total
flow stays positive at all scales no back-bending is found.

Noteworthy, the back-bending phenomenon does not only occur for a
strict discontinuity in the fermion flow. For any other tested
regulator back-bending can be observed as soon as the fermionic flow
contribution decouples completely (tends to zero) from the total flow
at a finite scale $k>0$, albeit the back-coupling seems to be stronger
the more rapidly the decoupling happens.

In summary, the choice of the regulator function has a crucial
influence on the momentum structure of the loop integrals of the flow
equations. Thus, an incongruous choice can potentially cause
unphysical regulator effects. An extreme example is given by the
discussed flat regulator in LPA which, albeit optimized in the vacuum,
cancels all momentum dependence of the propagator. As demonstrated
above this leads at finite chemical potential to a back-bending and
negative entropy density beyond the chiral transition. The mass-like
regulator does not modify the momentum structure at all and leads to a
smooth Callan-Symanzik type flow which does not produce such regulator
artifacts.

The Wetterich equation explicitly allows for an arbitrary choice of
regulator functions that fulfill the criteria summarized in
Sec. \ref{sec:regulator-functions}, see also
App.~\ref{app:optimization}, The flow equation in LPA, however, keeps
the momentum structure of the two-point function
$\Gamma_k^{(2)}(p) = p^2 + m_k^2$ generically fixed such that the flow
cannot compensate for the particular choice of momentum dependence for
$R_k(p^2)$.  Consequently, contributions which are sensitive to the
momentum structure, as for example the fermion decoupling discussed
above, can lead to a strong scheme dependency.

However, for more involved truncations where for example generally
momentum dependent wave function renormalization $Z_k (p)$ are taken
into account such regulator effects are not expected to appear anymore.

\section{Summary and Conclusions}
\label{sec:summary}

The functional renormalization group method is a powerful
non-perturbative tool that has a broad variety of research fields.
A key ingredient of the flow equation is the regulator that suppresses
the infrared physics via an infrared cutoff.  The optimal choice of
the regulator function plays a major role in the quantitative
optimization of actual calculations. Recently, great improvements
could be achieved by the application of the principle of minimum
sensitivity within $O(N)$-models which demonstrate the convergences of
the derivative expansion to accurate and precise results
\cite{DePolsi:2022wyb}.

In addition to a suitable choice of the regulator, any functional
equation must be truncated for technical reasons in general to obtain
a finite system of equations that allows a numerical treatment.
Feasible truncations in QCD applications are often more limited than
in simpler model studies such that one relies on more general
optimization criteria \cite{Litim2000,Pawlowski:2005xe}. In
particular, the impact of a finite chemical potential on possible
truncation errors is essentially unknown. Especially, at low
temperatures and finite density, strange effects like the observed
back-bending of the chiral transition line in various low-energy
effective models \cite{Schaefer:2004en, Tripolt:2021jtp} and the
associated occurrence of negative entropy densities beyond the
transition \cite{Tripolt2017} could hint at the existence of large
truncation artifacts.

The focus of this work is the regulator scheme dependence of
functional renormalization group equations at finite density. General
considerations on a reasonable choice of regulator including
convergence properties in particular at finite density are given and
confronted to various known optimization criteria in the literature.
As an application the chiral phase structure of the quark-meson model
at low temperatures has been calculated with three different
regulators in local potential approximation.  Within this
approximation we have found clear evidence that the back-bending of
the transition line and the odd appearance of a
negative entropy density are related to the choice of the regulator
function.

For momentum-independent Callan-Symanzik type regulators the chiral
phase transition is a smooth crossover for physical pion masses and
the transition line hits the $\mu$-axis perpendicularly. No negative
entropy densities are observed in contrast to the familiar
back-bending scenario with the optimized flat regulator.

This essentially allows for two different interpretations: firstly,
the Callan-Symanzik regulator, which seems to be an sub-optimal
regulator in critical exponents evaluations \cite{Litim:2002cf}, is
just incapable of resolving some of the physical intricacies leading
thus to the back-bending phenomenon in the phase structure, or
secondly, the back-bending is an actual unphysical artifact induced
by the nontrivial momentum structure of the flat regulator in local
potential approximation.

A detailed investigation of the renormalization group flow reveals
that the back-bending is induced at scales around the Fermi
surface. For the flat regulator, it appears as a discontinuity in the
fermion flow that is smeared out at finite temperature. This smearing
feeds back into the bosonic flow and leads to a large temperature
sensitivity which can eventually be observed in the curvature of the
transition line. For Callan-Symanzik type regulators, the Fermi
distributions are fully integrated out in the Wetterich loops and
therefore lead only to small, thermodynamically sensible
modifications. Moreover, at the level of the LPA the momentum
structure of the loop integral is fully determined by the choice of
regulator function because higher orders of momenta, such as of
$\mathcal{O}(p^4)$, are neglected in the effective action. In total,
these findings lead to the conclusion that the back-bending phenomenon
found in local potential approximation is a non-physical artefact
induced by the specific choice of the shape function.

We remark that the absence of the above mentioned truncation artifacts
for Callan-Symanzik regulators does not imply a full and sufficient
convergence meaning that the obtained results could still be quite
insufficient from a more quantitative viewpoint. One idea to
circumvent this possible constraint is the simultaneous combination of
regulator functions such that the accuracy can be improved and
additionally avoids the back-bending issue. Of course, when no
truncation is made at all, a full solution for the effective average
action does not depend on the choice of regulator, whereas any sort of
truncations introduces a spurious dependence on it. On a more advanced
level, by going beyond the LPA a more complex momentum structure of the
effective action might remedy the problem regardless of the regulator
function.  For example we expect that the inclusion of higher
momentum-dependent wavefunction renormalizations will certainly affect
such regulator effects but it is still an open issue how large the
truncation artifacts of higher truncation orders are.

\subsection*{Acknowledgments}

We thank Jan Pawlowski and Fabian Rennecke for interesting and
enlightening discussions.  We acknowledge support by the Helmholtz
Graduate School for Hadron and Ion Research for FAIR, the GSI
Helmholtzzentrum für Schwerionenforschung and the BMBF
under Contract No. 05P18RGFCA. KO acknowledges funding by the German
Academic Scholarship Foundation and BJS by the Deutsche
Forschungsgemeinschaft (DFG) through the grant CRC-TR 211
“Strong-interaction matter under extreme conditions”.

\appendix

\section{General Optimization Criteria}
\label{app:optimization}

Finding an \textsl{optimal} shape function $r(y)$ in the sense that
the truncation error is minimized, i.e., that for a given set of
observables $\{O_n\}$ their physical values are approached as closely
as possible, is a non-trivial task. So far, three different criteria
are found in the literature of which a brief summary shall be given in
this appendix.

The principle of minimum sensitivity (PMS) \cite{Stevenson:1981vj,
  *Ball:1994ji}, first applied in perturbation theory, aims at finding
solutions that are least sensitive to variations in the regularization
scheme. In practice, this usually entails a full computation of the RG
flow of the given set of observables $\{ O_n \}$ under variation of
one or more parameters of a parameterized shape function. A coinciding
extremum for all $O_n$ is searched for. However, such calculations are
computationally expensive and solutions are not necessarily unique or
exist at all, see e.g.~\cite{Pawlowski:2005xe} for a
discussion. However, despite those drawbacks this criterion has been
successfully applied in the past to, for example, the accurate
determination of critical exponents of the three-dimensional Ising
model \cite{Canet:2002gs, Canet:2004xe}.  Recently, successive
improvements have been made with the application of the PMS to fix the
regulator dependence which establishes the convergence of the
derivative expansion with great precision and accuracy
\cite{DePolsi:2022wyb, DePolsi:2021cmi, *DePolsi:2020pjk}.

An observable-independent criterion that provides very precise values
for critical exponents \cite{Litim:2002cf} has been put forward in
\cite{Litim2000}. It is based on the idea that regulators which
maximize the gap in the massless inverse dimensionless propagator
$P^2(y)$, \eqref{eq:inverse_propagator}, yield the greatest stability
of the flow and the quickest approach towards the physics in the IR:

\begin{equation}
\label{eq:optimal_gap}
C_\mathrm{opt} := \max_{R} \left(\min_{y \geq 0} P^2(y) \right) \ .
\end{equation}

One reasoning for this assumption is that for the largest gap an
expansion of the flow in inverse powers of $P^2$ leads to a most rapid
convergence of the series, see \cite{Litim:2001fd} for further
details. All regulators that fulfill \eqref{eq:optimal_gap}, further
need to obey a normalization condition since otherwise
$C_\mathrm{opt}$ could be made arbitrarily large. The usual choice is
to set
\begin{equation}
\label{eq:3}
R_k(y_0k^2) = y_0k^2
\end{equation}
or, equivalently, to fix $r(y_0) = 1$ for some positive $y_0 > 0$
\cite{Litim2000}. A more general form $r(y_0) = c$ with any finite
positive $c > 0$ is also possible \cite{Pawlowski:2005xe}. 

To ensure that $y_0$ exists and is unique, we restrict ourselves to
continuous, strictly monotonously decreasing shape functions $r(y)$ in
the following. This kind of normalization is closely related to the
introduction of an effective RG scale $k_\mathrm{eff}$: for a given
$r(y)$ with normalization $r(y_0)=1$ one can define a family of shape
functions
\begin{equation}
r^\lambda(y) := r(y/\lambda^2)
\end{equation}
where $r^\lambda$ is obtained from the original shape function $r$ by
a rescaling $k \rightarrow \lambda k$ with $\lambda > 0$. Clearly,
this shift in the effective RG scale leads to the same trajectory in
theory space, corresponding just to a different parameterization of
the effective action $\Gamma^\lambda_{k} \equiv \Gamma_{\lambda k}$
where $\Gamma^\lambda$ is the effective action obtained with
$r^\lambda$. If we now \textsl{set} $k_\mathrm{eff}^2 := y_0 k^2$, the
rescaling shifts $k_\mathrm{eff} \rightarrow \lambda k_\mathrm{eff}$
and $r^\lambda$ obtains a different normalization
$r^\lambda(\lambda^2 y_0) = 1$. Hence, for monotonous shape functions
the normalization singles out exactly one regulator from each family
of equivalent regulators that differ only by a constant multiplicative
shift in $k_\mathrm{eff}$. The optimization criterion
\eqref{eq:optimal_gap} thus only compares regulators with the same
$k_\mathrm{eff}$ which depends on $k$, i.e. $k_{\mathrm{eff}}(k)$. The
optimal gap is attained when $P^2(y)$ has its minimum at $y_0$. With
the normalization $r(y_0)=1$ in \eqref{eq:inverse_propagator} the
optimal gap in LPA and for vanishing fields is thus $C_\mathrm{opt} = 2 y_0$ according to this criterion
\cite{Litim2001}.

A popular choice of an optimal regulator is the flat (or Litim)
regulator \cite{Litim2001}
\begin{equation}
  \label{eq:4}
r_\mathrm{flat}(y) = \left(\frac{1}{y} - 1 \right) \, \Theta \left(1-y\right)  \ .
\end{equation}
This is not the unique solution to \eqref{eq:optimal_gap}.  Many more
regulator shape functions that fulfill the optimization criterion can
be found in the literature. Oftentimes, they can be obtained from
generalized regulator classes like the compactly supported smooth
(CSS) regulators \cite{Nandori:2012tc}.  However, the flat regulator
(\ref{eq:4}) is special in the sense that the shape function is
optimal for \textit{any} arbitrary normalization point $c$, i.e.,
$r(y_0) = c$ with $y_0 = 1/(c+1)$.

The special role of the flat regulator in LPA is further confirmed by
a third functional optimization criterion developed in
\cite{Pawlowski:2005xe, Pawlowski2015}. Is is also grounded on a stability
assumption for  correlation functions that should be
insensitive to local variations of the regulator at a fixed physical
cutoff scale $k_\mathrm{phys}$. The physical cutoff scale
$k_\mathrm{phys}$ is here given  by the gap of the inverse
propagator.

The condition can be related to the minimization of the kernel of the
flow operator $\partial_t$ that minimizes the total length of the flow
trajectory in the theory space, see  \cite{Pawlowski2015} for further
details. In LPA and for a single scalar field the criterion reduces to
a bounded
dimensionless shape function for all momenta,
\begin{equation}
\label{eq:opt_pawlowski}
r_\mathrm{opt}(y) \leq r \ , \quad \forall r,y \ ,
\end{equation}
with the normalization condition
\begin{equation}
\label{eq:normalization_pawlowski}
\min_{y \geq 0} P^2(y) = k_\mathrm{phys}^2/k^2 \ .
\end{equation}
In contrast to the previous criterion \eqref{eq:optimal_gap} where the
gap is maximized for shape functions intersecting at a common point,
this criterion compares regulators leading to the same gap and chooses
the one that maximizes the propagator over the whole spectrum. For the
\textit{special} choice $k_\mathrm{phys} = k$, the flat
regulator is the unique solution to the criterion
\eqref{eq:opt_pawlowski}.

We close this recapitulation of optimization criteria with two
remarks: the mentioned optimization criteria do not change when the
truncation of the effective action is improved beyond LPA by taking
momentum-independent wave function renormalization $Z_k$ into account,
often denoted as LPA' in the literature. For example, augmenting
\eqref{eq:15} with the wavefunction renormalization,
\begin{equation}
  \label{eq:5}
R_k(p^2) = Z_k \, p^2 \, r(y) \ ,
\end{equation}
implies the propagator modification $P^2 \to Z_k P^2$ and allows for a
systematic inclusion of higher derivative operators in \eqref{eq:1}.

Secondly, the previous arguments and criteria can be straightforwardly
adapted to fermions. To preserve, e.g., chiral and gauge symmetries, a
regulator resembling a kinetic term, cf.~\eqref{eq:15}, can be chosen as
\begin{equation}
\label{eq:fermi_regulator}
R_k^F(p) = Z_k \mathrm{i}\slashed{p} \, r^F(y) \ .
\end{equation}
The fermionic analogon to
\eqref{eq:inverse_propagator} (for $Z_k = 1$) follows from the inverse
propagator
\cite{Litim2001}
\begin{equation}
P_F^2(y) = y [1+r^F(y)]^2 \ ,
\end{equation}
such that the choice
\begin{equation}
\label{eq:fermi_regulator_shape}
r^F(y) = \sqrt{1+r(y)} - 1
\end{equation}
leads to the same effective regulator scheme. This yields the
fermionic version of the flat regulator shape function
\begin{equation}
\label{eq:fermi_flat_regulator}
r_\mathrm{flat}^F(y) = \left(\sqrt{\frac{1}{y}} -1 \right) \Theta(1-y) \ .
\end{equation}

\section{Approximate Flows in the Chirally Symmetric Regime}
\label{app:approximate_flows}
In \cite{Berges:1997eu} it was observed that the the two-flavor quark-meson model exhibits an approximate partial IR fixed point behavior in the chirally symmetric regime. This can be taken advantage of to constrain the effective potential at the chiral symmetry breaking scale $k_\chi$. We detail here the technical derivation of the fixed point values used in Sec. \ref{sec:parameter-fixing} in a general, regulator-independent way. This follows closely the arguments made in \cite{Berges:1997eu} but generalizes them to regulators that incorporate an additional scale, such as the mass-like regulators \eqref{eq:4dstepreg} and \eqref{eq:3dstepreg} which depend on $k_\phi$. The chain of argument relies on the consideration of vacuum flows in LPA', i.e., with running wavefunction renormalizations and a running Yukawa coupling. At large RG scales $k > k_\chi$, fluctuations are dominated by the purely fermionic contributions. Picking up the definitions for $\tilde{\rho}$ and $u_k(\tilde{\rho})$, \eqref{eq:dimensionless_chiral_invariant} and \eqref{eq:dimensionless_ren_potential}, the approximate flow of the dimensionless potential at fixed $\tilde{\rho}$ becomes
\begin{equation}
\label{eq:running_dimensionless_potential}
\partial_t u_t(\tilde{\rho}) = -4 u_t + (2+\eta_{\phi,t}) \, \tilde{\rho} \, u_t'(\tilde{\rho}) - \frac{N_cN_f}{4\pi^2} l_{0,t}^F(\tilde{m}_{\psi,t}^2) \ .
\end{equation}
To simplify the notation in the upcoming discussion, an explicit scale dependence is now expressed by the RG time $t=\ln(k/k_\phi)$ instead of the corresponding dimensionful scale $k$ in the lower index. The fermion loop is expressed in terms of the threshold function which reads for 4d regulators
\begin{equation}
l_{0,t}^{F,\mathrm{4d}}(\tilde{m}_{\psi,t}^2) = \int_0^\infty \mathrm{d}y \, y^2 \, \frac{\left[1+r^F_t(y)\right] \, \partial_t r^F_t(y)}{y \left[1+r_t^F(y)\right]^2 + \tilde{m}_{\psi,t}^2} 
\end{equation}
and for 3d regulators
\begin{equation}
l_{0,t}^{F,\mathrm{3d}}(\tilde{m}_{\psi,t}^2) = 2 \int_0^\infty \mathrm{d}x \, x^{3/2} \, \frac{\left[1+r_t^F(x)\right]  \partial_t r_t^F(x)}{\sqrt{x\left[1+r_t^F(x)\right]^2 + \tilde{m}_{\psi,t}^2}} \ .
\end{equation}
The dimensionless quark mass is given by
\begin{equation}
\tilde{m}_{\psi,t}^2 = \frac{\bar{g}_t^2}{2} \tilde{\rho} \ .
\end{equation}
Note the factors in the definitions of the threshold functions have been chosen in agreement with \cite{Berges:1997eu} for better comparability. We allow for an explicit scale dependence of $r_t^F$ beyond that of its argument $y=p^2/k^2$ or $x=\bm{p}^2/k^2$, as denoted by the index $t$. Such a dependence exists, for example, for the mass-like regulators via the dimensionless UV scale $\tilde{k}_\phi := k_\phi/k = \E^{-t}$. If the shape function does not possess an explicit scale dependence, the threshold functions $l^F_n(\tilde{m}_{\psi,t}^2)$ only depend on $t$ implicitly via the quark-mass argument. By definition, the higher-order threshold functions $l_{n,t}^F$ are related to $l_{0,t}^F$ via \cite{Berges:1997eu}
\begin{equation}
l_{n,t}^F(\tilde{m}_{\psi,t}^2) := \frac{(-1)^n}{(n-1)!} \left(\frac{\mathrm{d}}{\mathrm{d}\tilde{m}_{\psi,t}^2}\right)^n l_{0,t}^F(\tilde{m}_{\psi,t}^2) \ .
\end{equation}
The quark anomalous dimension generally vanishes in this approximation and the meson anomalous dimension can be written
\begin{equation}
\eta_{\phi,t} = \frac{N_c N_f}{16 \pi^2} \, \bar{g}_k^2 \, \kappa^F_t \ .
\end{equation}
$\kappa^F_t$ denotes the purely fermionic part of the corresponding threshold function evaluated at $\tilde{\rho}=0$. We do not give an explicit expression as it is not required. The flow of the Yukawa coupling is solely fed by its renormalization,
\begin{equation}
\partial_t \bar{g}_t^2 = \eta_{\phi,t} \, \bar{g}_t^2 \ ,
\end{equation}
and has the solution
\begin{equation}
\bar{g}_t^2 = \frac{\bar{g}_0^2}{1-\frac{N_cN_f}{16 \pi^2} \bar{g}_0^2 \int_0^t \mathrm{d}s \, \kappa^F_s}
\end{equation}
where $g_0$ is the initial value at $t=0$. Expanding $u_t(\tilde{\rho})$ in a power series, see \eqref{eq:power_expansion_uk}, flow equations for the coefficients $u_t^{(n)}(0)$ can be inferred from \eqref{eq:running_dimensionless_potential} and similar exact solutions. For $n=2$, one finds
\begin{equation}
\label{eq:solution_fixedpoint_2}
\frac{u^{(2)}_t(0)}{\bar{g}^2_t} = \frac{\frac{u_0^{(2)}(0)}{\bar{g}_0^2}-\frac{N_cN_f}{16 \pi^2}\bar{g}_0^2\int_0^t \mathrm{d}s \, l^F_{2,s}(0)}{1-\frac{N_cN_f}{16 \pi^2}\bar{g}_0^2 \int_0^t \mathrm{d}s \, \kappa^F_s} \ .
\end{equation}
Note that for any regulator shape function $r^F$ without explicit $t$ dependence, one obtains the simple relations $\kappa^F=l_2^F(0) \equiv 1$ \cite{Berges:1997eu}. In this case, the solution simplifies to
\begin{equation}
\label{eq:solution_fixedpoint_2_simplified}
\frac{u^{(2)}_t(0)}{\bar{g}^2_t} = 1 - \frac{1 - \frac{u_0^{(2)}(0)}{\bar{g}_0^2}}{1-\frac{N_cN_f}{16 \pi^2}\bar{g}_0^2 \, t}
\end{equation}
and for $t \rightarrow -\infty$ approaches the infrared fixed point
\begin{equation}
\left. \frac{u^{(2)}_t(0)}{\bar{g}^2_t} \right\vert_* = 1 \ .
\end{equation}
For the two mass-like regulators, the same infrared value is approached even though they require the more complex solution \eqref{eq:solution_fixedpoint_2}: At large negative RG times t, the UV cutoff parameter $\tilde{k}_\phi = \E^{-t}$ diverges quickly and both $\kappa_t^F$ and $l^F_{2,t}(0)$ become effectively scale-independent, tending to unity. Thus, the leading contributions to the integrals in \eqref{eq:solution_fixedpoint_2} behave like $t$ and all subleading terms vanish for $t \rightarrow - \infty$. A similar analysis works at all orders $n \geq 3$ where the solution for the expansion coefficients reads
\begin{align}
\begin{split}
\frac{u_t^{(n)}(0)}{\bar{g}_t^{2n}} &= \E^{2(n-2)t} \, \frac{u^{(n)}_0(0)}{\bar{g}_0^{2n}} - \frac{N_cN_f}{4 \pi^2} \, \frac{(-1)^n (n-1)!}{2^n} \, \times \\
& \hspace{2cm} \times \E^{2(n-2)t} \, \int_0^t \mathrm{d}s \, l_{n,s}^F(0) \, \E^{-2(n-2)s} \ .
\end{split}
\end{align}
In the case of scale-independent threshold functions $l^F_n(0)$, the integral can be solved trivially and the infrared fixed point is given by 
\begin{equation}
\label{eq:solution_fixedpoint_n_simplified}
\left. \frac{u_t^{(n)}(0)}{\bar{g}_t^{2n}} \right\vert_* = \frac{N_cN_f}{8 \pi^2} \frac{(-1)^n (n-1)!}{2^n (n-2)} \, l_n^F(0) \ .
\end{equation}
At these orders of the expansion, the threshold functions depend on the explicit choice of shape function, i.e., they yield different fixed points for different regulators. From similar arguments as above, it follows that the infrared-attractive points for the mass-like regulators are determined by inserting the asymptotic threshold functions $l^F_{n,t \rightarrow -\infty}(0)$ for $l_n^F(0)$ in \eqref{eq:solution_fixedpoint_n_simplified}.

\section{Pole Proximity of Vacuum Flows}
\label{app:pole_proximity}

In general, the non-perturbative flow equations are composed of
threshold functions that accommodate prospective singularities
governed by the sign of the potential derivatives.  A typical
phenomenon occurs for vacuum flows of quark-meson model truncations
(or similar theories) in LPA on a discretized $\sigma$-field grid: for
small $\sigma$-values the pion threshold function is the dominant one
in the IR and the pion mass $m_\pi^2 = 2 U_k'(\sigma^2)$ becomes
negative in the vicinity of the pole in the (Euclidean) propagator.
During the remaining IR evolution it follows closely along this
regulator-dependent pole.  This poses a significant numerical
challenge since small numerical deviations can easily hit
this pole.

In this appendix we estimate the proximity of the pion pole
analytically as follows:
The effective inverse pion propagator in LPA
\begin{equation}
  \label{eq:11}
\Gamma_{\pi, k}^{(2)}(p^2)+R_k(p^2) = k^2 P^2(p^2/k^2) + m_\pi^2 \ ,
\end{equation}
exhibits a pole in the momenta as soon as $m_\pi^2$ falls below a
certain negative threshold (assuming the positivity of the inverse
propagator $P^2$, \eqref{eq:inverse_propagator}).  The threshold is
determined by the negative of the massless propagator gap
\begin{equation}
\label{eq:pole_condition}
m^2_{\pi, \mathrm{thres}} = - k^2 \min_{y \geq 0} P^2(y) \ .
\end{equation}
This argument can straightforwardly be transferred to dimensionally
reduced regulators. For a 3d regulator the two-point function reads
accordingly
\begin{equation}
  \label{eq:12}
\Gamma_{\pi,k}^{(2)}(p_0, \bm{p}^2)+R_k(\bm{p}^2) = p_0^2 + k^2 P^2(\bm{p}^2/k^2) + m_\pi^2 \ ,
\end{equation}
such that the threshold value at the minimal $p^2_0=0$ is still given by
\eqref{eq:pole_condition}.

For a 3d flat regulator the minimum of the massless inverse propagator
is at $P^2 (x) = 1$ and the pole for vanishing $\sigma$-field is
located at $U_k' = - k^2/2$, cf.  \eqref{eq:vacuum_flow_litim3d}. For
the dimensionless variable $\tilde{u}'_k := 2 U_k'/k^2$ the pion
propagator pole is shifted to $\tilde{u}_k' = -1$ with the
corresponding flow equation evaluated at $\sigma=0$
\begin{align}
\begin{split}
\label{eq:dtuktilde_flat}
  \partial_t \tilde{u}'_k &= 2 (\partial_t U_k')/k^2 - 2 \tilde{u}'_k \\ 
  &= \frac{1}{\pi^2} \left[- \frac{U_k''}{(1+\tilde{u}'_k)^{3/2}} +
    \frac{\nu}{12} \left(\frac{g}{2}\right)^2 \right] - 2 \tilde{u}'_k
  \ . 
\end{split}
\end{align}
A similar analysis for the 3d mass-like regulator,
cf. ~\eqref{eq:3d_vacuum_flow}, yields the flow equation (again
evaluated at $\sigma=0$)
\begin{align}
	\begin{split}
		\label{eq:dtuktilde_mass}
		\partial_t \tilde{u}'_k = & \frac{3U_k''}{\pi^2}
                \left[ \frac{\tilde{k}_\phi}{
                    \sqrt{\tilde{k}^2_{\phi} +1 + \tilde{u}'_k }}
                  - \text{artanh} \left( \frac{\tilde{k}_\phi}{
                      \sqrt{\tilde{k}^2_{\phi} +1 + \tilde{u}'_k } }
                  \right) \right] \\ 
		- \frac{\nu}{4 \pi^2} &\left(\frac{g}{2}\right)^2
                \left[ \frac{\tilde{k}_\phi}{ \sqrt{ \tilde{k}^2_{\phi}
                      +1 }} - \text{artanh} \left(
                    \frac{\tilde{k}_\phi}{ \sqrt{ \tilde{k}^2_{\phi}
                      +1 } } \right) \right] - 2
                \tilde{u}'_k
	\end{split}
\end{align}
that additionally depends on the dimensionless parameter
$\tilde{k}_\phi := k_\phi/k \geq 1$ which increases during the IR
evolution.
 
\begin{figure}
  \centering
  \includegraphics[width=\onefig]{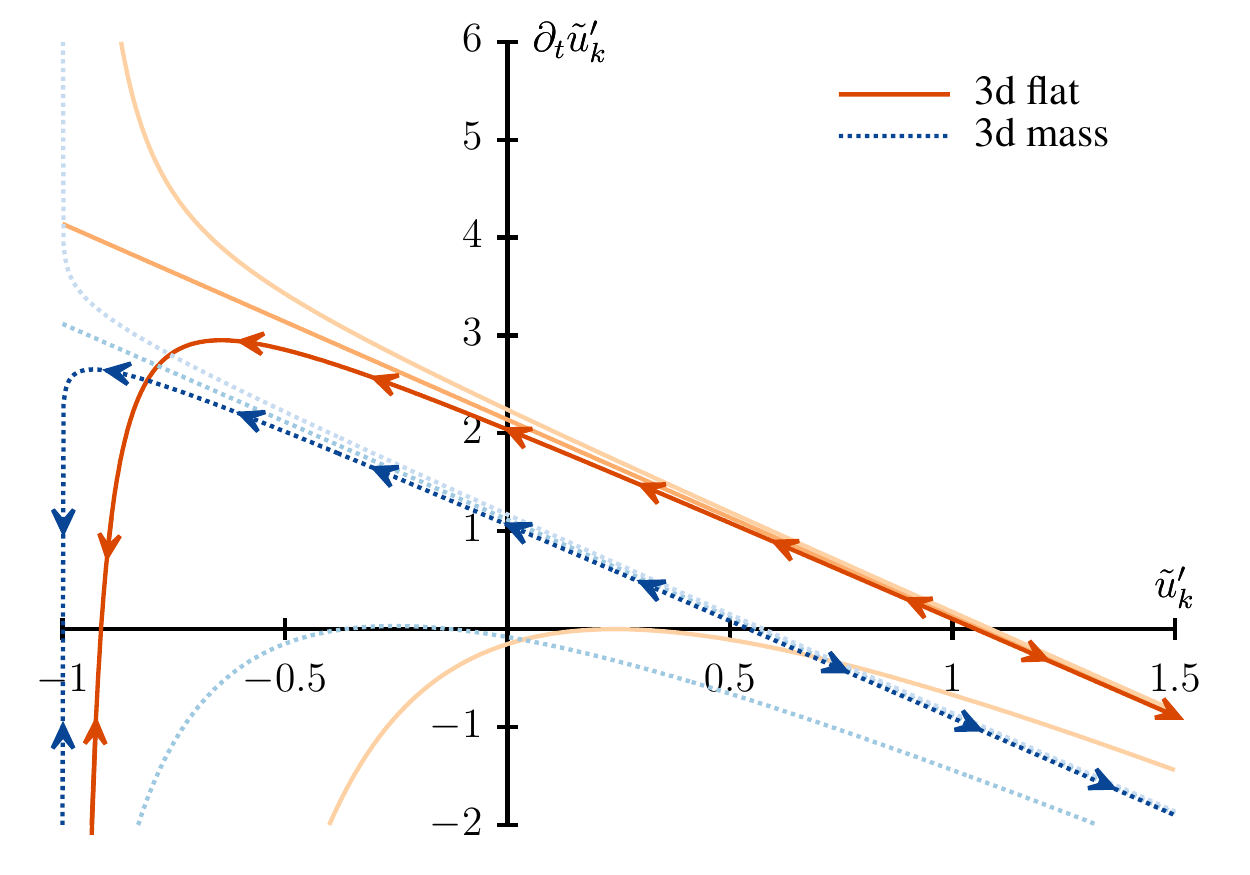} 
  \caption{\label{fig:fixed_points} Flow of $\tilde{u}'_k$ evaluated
    at $\sigma=0$ for the two 3d regulators,
    cf. Eqs. (\ref{eq:dtuktilde_flat}) and
    (\ref{eq:dtuktilde_mass}). $U_k'' = (-1,0,1, 22.6)$ (from top to
    bottom) and $\tilde{k}_\phi = 1$ for the mass-like regulator.}
\end{figure}

Both equations are displayed in \Fig{fig:fixed_points} where the flows
$\partial_t \tilde{u}^{\prime}_k$ are given as a function of $\tilde{u}'_k$
for four different $U''_k = (-1,0,1,22.6)$ (from top to bottom). The
solid orange lines are the results for the 3d flat regulator and the
dashed blue lines for the 3d mass-like regulator (for
$\tilde{k}_{\phi}=1$). The parameters in both quark flow contributions
are kept fixed to $g=6.5$ and $\nu=24$.

Generically, the structure of the flow equation for all discussed
regulators is very similar at $\sigma=0$ : For $U_k''>0$ but below a
certain positive value $U^{* \prime\prime}_k$ two stationary points
defined by $\partial_t \tilde{u}'_k=0$ appear and for $U_k''<0$ only
one stationary point (in the figure: the one on the right side where
$\tilde{u}'_k>0$) survives. Note that these points are not actually
fixed points since their location depends on $U_k''$ and changes during
the flow. For increasing $U_k''$ both stationary points come closer to
each other and degenerate at $U^{* \prime\prime}_k$
($U^{* \prime\prime}_k \sim 22.6$ for the 3d flat regulator and
$U^{* \prime\prime}_k \sim 23.1$ for the 3d mass-like regulator). For
larger values they disappear completely.

Since the flow equations are integrated in negative $t$-direction,
$\tilde{u}'_k$ increases for $\partial_t \tilde{u}'_k < 0$ and
decreases for $\partial_t \tilde{u}'_k > 0$. As a
consequence, the right stationary point is a repulsive point (it has a
negative slope) while the left point in the vicinity of the pole is an
attractive one with a positive slope. The flow pattern of two
exemplary curves are indicated by arrows in \Fig{fig:fixed_points}.

Therefore, for $U''_k \in \left] 0, U^{* \prime\prime}_k \right[$ and
for values of $\tilde{u}'_k$ smaller than the repulsive point, the
derivative is always pushed towards the attractive left point and thus
never runs into the pole. It might be that the flow oscillates around
this attractive point towards the IR which would aggravate its
numerical treatment. For a negative $U_k''$ only the right repulsive
point exists in the flow pattern but this case would lead to
a flow directly into the pole for $\tilde{u}'_k$
  values smaller than this point. For $\tilde{u}'_k$ values larger
  than the repulsive point a permanent flow towards increasingly
  positive values emerges.  Similar, for sufficiently large
$U_k'' > U^{* \prime\prime}_k$ no stationary points exist anymore and
the flow is always driven to positive values, avoiding a chiral
symmetry breaking since $\tilde{u}'_k$ becomes increasingly large.

Due to the convexity of the Wetterich flow equation the eventually IR
evolved potential is also convex such that the physical relevant case
is the one where the second potential derivative
$U_k''$ tends to zero. This pushes, according to
\eqref{eq:dtuktilde_flat}, the stationary point around the pole even
closer to it.  In order to estimate the pole proximity we introduce
the quantity $\delta u$ through $\tilde{u}'_k = -1 + \delta
u$. Multiplying both sides of \eqref{eq:dtuktilde_flat} with
$\delta u^{3/2}$ and setting $\partial_t \tilde{u}'_k = 0$ we find to
lowest order in $\delta u$
\begin{equation}
\label{eq:deltau_flat}
\delta u \approx \left( \frac{U_k''}{2 \pi^2 + \frac{\nu}{12}
    \left(\frac{g}{2}\right)^2}\right)^{2/3} \ . 
\end{equation}
For $U_k''=1$ this yields approximately $\delta u \approx 0.0843$
and  is in agreement with an error of less than 1$\%$ in the previous
numerical calculated pole $\tilde{u}{'}_k^{(0)}$.

A similar treatment of the flow equation for the 3d mass-like
regulator yields a comparable pole proximity. Explicitly, setting
$\partial_t \tilde{u}'_k = 0$, the poles in \eqref{eq:dtuktilde_mass}
can be eliminated by exponentiating again both sides with the result to
lowest order

\begin{equation}
  \label{eq:13}
  \delta u \approx 4 \tilde k_\phi^2 \, \exp \left[
    -2- \frac{8 \pi^2}{12 U_k''}
    \left( \left. \partial_t \tilde{u}'_k \right|_\psi + 2 \right) \right]
      \ ,
\end{equation}
where $\left.\partial_t \tilde{u}'_k \right|_\psi$ is the $U''_k$ independent (but $k_\phi$  dependent) fermionic part of the flow for $\tilde{u}'_k$, see \eqref{eq:dtuktilde_mass}.

Already for $U_k''=1$ and for $\tilde{k}_\phi = 1$ one finds
$\delta u \approx 6.619 \times 10^{-10}$ being significantly closer to
zero (and much smaller in comparison to the flat regulator
proximity). This is nicely visible in \Fig{fig:fixed_points}.

A comparison of both estimates reveals that for the mass-like
regulator the proper numerical treatment of the corresponding flow
equations is much more involved. Furthermore, the estimate is further
suppressed exponentially with decreasing $U_k''$ compared to the
power-law suppression for the flat regulator. However, the situation
does not change significantly in the 4d regulator case. For
completeness,  the flow equation with a 4d mass-like regulator 
\begin{align}
  \label{eq:14}
\begin{split}
\partial_t \tilde{u}'_k = &- \frac{3U_k''}{2 \pi^2} \left[- \frac{1}{1+
    \frac{1+\tilde{u}'_k}{\tilde{k}_\phi^2}} + \ln \left(1 +
    \frac{\tilde{k}_\phi^2}{1+\tilde{u}'_k}\right) \right] \\ 
&+ \frac{\nu}{8 \pi^2} \left(\frac{g}{2}\right)^2 \left[ - \frac{1}{1+
    \tilde{k}_\phi^{-2}}+ \ln (1+\tilde{k}_\phi^2)\right] - 2
\tilde{u}'_k 
\end{split}
\end{align}
leads to the pole proximity
\begin{equation}
\delta u \approx \frac{\tilde{k}_\phi^2}{\exp \left[\frac{2\pi^2}{3U_k''}
		\left( 2 + \left.\partial_t \tilde{u}'_k \right|_\psi \right) +1 \right]-1}   
\end{equation}
that yields $\delta u \approx 1.198 \times 10^{-8}$ for the same
parameters.

Already in \cite{Litim2000} it was shown that an optimized regulator
according to the criterion \eqref{eq:optimal_gap} pushes the
propagator poles as far as possible down on the negative $U_k'$-axis
and it was speculated that such regulators, in particular the flat
regulator, thus lead to the smoothest and numerically most stable
flow. For the mass-like regulator, the pion pole proximity seems to be
particularly grave as demonstrated in \Fig{fig:fixed_points}. A
numerical solution for the vacuum flow, as argued above, is with
standard methods not possible.

Note that in this work a numerical solution of this issue was not
necessary due to the Silver Blaze property at $T=0$ and
$\mu \lesssim \mu_c$.  For this parameter regime the potential is only
modified around small field values. Around the vacuum expectation
value the potential remains unchanged and allows for a proper
determination of the vacuum masses and condensate.

For future applications the analytical estimates for the stationary
points in the flows might be an additional useful reference to
stabilize the numerical setup

\section{Numerical Implementation}
\label{app:numimpl}

In this appendix the numerical procedure for solving the flow
equations in LPA is provided. Generally, a flow equation for the
effective potential is a partial differential equation (PDE) for two
independent variables $t$ and $\sigma$. They feature a first-order
derivative of the potential with respect to the logarithmic RG scale
$t = \ln k$ and a first- and second-order potential derivative with
respect to the square of the radial $\sigma$-mode in field space,
cf.~\eqref{eq:6}, which turn them into coupled highly non-linear
equations. Traditionally, these equations are solved on (equidistant)
grids in the field variable $\sigma$ or $\sigma^2$, with field
derivatives obtained from finite differences, coupled Taylor-grid
approximations \cite{Adams1995, *Bohr:2000gp}, or cubic splines
\cite{Mitter:2013fxa, *Resch:2017vjs}. Global approaches with
pseudo-spectral methods have also been employed
\cite{Borchardt:2016pif}. Recently, it has been shown that PDEs of the
Wetterich equation--type in LPA can be recast into another shape:
using the flow for the first potential derivative $\partial_t U_k'$ a
conservative form with distinct convective and diffusive fluxes can be
constructed \cite{Grossi:2019urj}.  This admits a modern treatment
within a hydrodynamic framework, utilizing finite volume methods
\cite{Koenigstein:2021syz,*Koenigstein:2021rxj,*Steil:2021cbu,*Stoll:2021ori}
or more advanced setups like discontinuous Galerkin methods
\cite{Grossi:2019urj, Grossi:2021ksl}. Especially, shocks occurring in
flows with a flat regulator caused by the discontinuity at the Fermi
surface can be resolved in great detail within such a novel
framework. Recently, in a first detailed study of the phase diagram of
the quark-meson model with discontinuous Galerkin methods
\cite{Grossi:2021ksl} the back-bending behavior of the chiral
transition line at finite densities has also been observed which
demonstrates that the back-bending is not an artefact of the numerical
implementation to solve the flow equations. We therefore retain a more
well-tried, computationally less expensive setup with a simpler
implementation based on cubic splines over an equidistantly
distributed grid in the $\sigma$-field space. The two missing boundary
conditions for the spline are obtained by fixing the first derivative
at the left- and rightmost points of the interval via a three-point
finite difference stencil. The chosen interval is
$\sigma \in [0,170 \, \mathrm{MeV}]$, and in most cases $n=40$ grid
points are used, with up to 80 points for the computation of the
crossover lines. All numerical results obtained in this way have been
cross-checked with the Taylor-grid method as outlined in
\cite{Adams1995}. For the solution of the coupled set of ordinary
differential equations (ODEs) an explicit higher-order Runge-Kutta
type ODE stepper with adaptive stepsize has been implemented
\cite{DORMAND198019}.

\bibliography{regdep}

\end{document}